\documentclass[preprint,1p]{elsarticle}
\usepackage{amsmath}
\usepackage{graphicx}
\usepackage{amssymb}
\usepackage{bm}
\newcommand{\beq}{\begin{equation}}
\newcommand{\eeq}{\end{equation}}
\newcommand{\beqa}{\begin{eqnarray}}
\newcommand{\eeqa}{\end{eqnarray}}

\journal{Nuclear Physics A}

\begin{document}

\begin{frontmatter}

\title{Transport properties and Langevin dynamics of heavy quarks and quarkonia
  in the Quark Gluon Plasma} 

\author[dep,infn]{A. Beraudo}

\author[infn]{A. De Pace}

\author[dep,infn]{W.M. Alberico}

\author[dep,infn]{A. Molinari} 

\address[dep]{Dipartimento di Fisica Teorica dell'Universit\`a di Torino,\\
via P.Giuria 1, I-10125 Torino, Italy}
\address[infn]{Istituto Nazionale di Fisica Nucleare, Sezione di Torino, \\ 
  via P.Giuria 1, I-10125 Torino, Italy}

\begin{abstract} 
Quark Gluon Plasma transport coefficients for heavy quarks and $Q\overline{Q}$
pairs are computed through an extension of the results obtained for a hot QED 
plasma by describing the heavy-quark propagation in the eikonal approximation
and by weighting the gauge field configurations with the Hard Thermal Loop 
effective action. It is shown that such a model allows to correctly reproduce,
at leading logarithmic accuracy, the results obtained by other independent
approaches. 
The results are then inserted into a relativistic Langevin equation allowing
to follow the evolution of the heavy-quark momentum spectra. Our numerical
findings are also compared with the ones obtained in a strongly-coupled
scenario, namely with the transport coefficients predicted (though with some
limitations and ambiguities) by the AdS/CFT correspondence. 
\end{abstract}

\end{frontmatter}

\section{Introduction}

Ultra-relativistic heavy-ion experiments at the Relativistic Heavy Ion Collider
(RHIC) and (in the near future) at the Large Hadron Collider (LHC)
aim at reproducing, in a small region of space-time, the conditions of the
primordial universe, with such a large energy-density to allow the onset
of the deconfined phase of QCD. Through high-energy nucleus-nucleus collisions
one should be able to explore the QCD phase-diagram in the region of high
temperature and (almost) vanishing baryon density. 

However the fireball of quarks and gluons possibly produced in such collisions
expands and cools, so that the only strongly-interacting particles reaching the
detectors are again color-singlet hadrons. 
Hence the need to extract information on what occurred before hadronization:
whether a thermalized Quark Gluon Plasma (QGP) was produced and, if so, which
are its properties. 

Heavy quarks (charm and bottom) and quarkonia are well suited for this
purpose. Due to their large mass they are produced in the very early 
stages of the collision and, before giving rise to experimentally detectable
signals from their decays into hadrons and/or leptons, they have to cross the
hot (possibly) deconfined region. By comparing the data collected in
nucleus-nucleus collisions with the ones coming from benchmark experiments,
like proton-proton or proton-nucleus collisions, in which one does not expect
QGP formation, one eventually can prove the formation of a totally new kind of
medium. 
In particular, a lot of interest on quarkonia was triggered by the seminal paper
by Matsui and Satz \cite{Mat} proposing the anomalous suppression of the
$J/\Psi$ in nucleus-nucleus collisions as an unambiguous signature of
deconfinement. 

For what concerns heavy quarks --- giving rise to D and B mesons at
hadronization --- they can provide useful insight into the transport properties
of the medium they cross. Due to their large mass they require a much longer
time (by a factor $M/T$) than the light particles to thermalize; naively one
expects that in an expanding medium with a quite short life-time (of order of 
$5-10$ fm/c) they can hardly approach thermal equilibrium. If, on the other
hand, one found them to follow the flow of the fluid, this would be a signal of
large friction and momentum-diffusion coefficients, whose origin would need an
explanation. 

RHIC data led to a description of the matter produced in the Au-Au collisions
as a strongly-coupled QGP (sQGP). In particular the radial \cite{rad1,rad2} and
elliptic (in semi-central collisions) flow \cite{v21,v22} observed in the
$p_T$-spectra of detected hadrons suggested a picture of the QGP in the regime
of temperatures accessible at RHIC more similar to a fluid (hence with very
small mean-free-paths) --- with a collective expansion driven by pressure
gradients --- than to a weakly-interacting gas of quarks and gluons. 
Furthermore, the strong suppression (by a factor of $5$) of high-$p_T$ hadrons
\cite{RAA1,RAA2} with respects to elementary proton-proton collisions suggests
that the QGP produced at RHIC is a very opaque medium. 

Heavy quarks at RHIC are
studied through the electrons arising from the semi-leptonic decays of D and B
mesons. This makes it hard to draw definite conclusions on their flow and
energy loss, due in particular to the difficulty of disentangling the charm and
bottom contributions. However, recent measurements performed by the PHENIX
\cite{phenix} and STAR \cite{star} collaborations seem to support a scenario
with the charm quark following the flow of the medium and experiencing a strong
energy loss, comparable to the one seen in  light hadron spectra. 
Different explanations where advocated to account for such an apparently rapid
thermalization, like the existence of resonant $Q\bar{q }$ states \cite{rapp1}
above $T_c$ or values of transport coefficients close to the bounds predicted
by the AdS/CFT correspondence \cite{yaf,sol1,sol2,gub,raja,horgyu}, claimed to
represent a good description of the sQGP for temperatures nearby $T_c$. 
See also Refs.~\cite{VHees,Gos1,Gos2,Gos3} for other approaches.

At LHC on the other hand one expects to produce a fireball with a larger
initial temperature (possibly $T_0\sim800$~MeV) and a longer
life-time. This could make weak-coupling calculations better justified. From
the experimental side \cite{ppr}, the possibility of reconstructing also
hadronic decays, like $D^0\to K^-\pi^+$, in Pb-Pb collisions will remove the
ambiguities encountered at RHIC. 

Two are the main issues addressed in this paper, both related to the transport
properties of charm and bottom quarks in a hot plasma. 
We first focus on the calculation of the transverse $\kappa_T(p)$ and
longitudinal $\kappa_L(p)$ momentum-diffusion coefficients. 
We then make use of the above results to follow the approach of heavy
quarks to thermal equilibrium through a numerical Langevin simulation. We
confine ourselves to the case of a static medium, for temperatures spanning a
range of experimental interest for LHC.

In order to calculate the coefficients $\kappa_T(p)$ and $\kappa_L(p)$ we
perform a generalization to the QCD case of an approach developed in
\cite{hq0,hq1} to describe heavy (static) ``quarks'' in a hot plasma of
electrons, positrons and photons. The latter represented in fact a system
sharing important features with the QGP, but for which the possibility of
employing in the calculations a Hard Thermal Loop (HTL) gaussian effective
action allowed to study in an easy way important medium effects. 
This approach makes it possible to
calculate also transport coefficients of $Q\overline{Q}$ pairs. This is
important, since measurements of $p_T$ spectra of quarkonia may allow to
discriminate different scenarios. One expects in fact that the measured
$J/\psi$'s in the final state should tend to follow the flow of the fireball, if
they came from the recombination of charm quarks free of floating around in the 
plasma phase. On the contrary, if they remained bound and sufficiently close in
space also in the deconfined environment, they would suffer less collisions in
the colored medium, as they would be seen mostly as a neutral object.

As already mentioned, we then insert the above transport coefficients into a
relativistic Langevin equation, allowing to follow the stochastic momentum
evolution of the heavy quarks in the QGP. 
Their initial momentum $p_0$ is taken of the order of the average transverse
momentum $\overline{p}_T$ of $c$ and $b$ quarks produced in nucleus-nucleus
collisions at RHIC and LHC. The latter can be, for instance, evaluated from a
sample of events generated by PYTHIA \cite{pythia} with a proper tuning of the
parameters.

Dealing with the relativistic Langevin equation, with a momentum dependent
diffusion term, is a non-trivial task. Actually, the latter belongs to the class
of stochastic differential equations, which represents an open field of
mathematical research, and requires a careful discretization procedure. In
particular, the drag coefficient has to be appropriately tuned in order to lead,
in the continuum limit, to a Fokker-Planck equation admitting the relativistic
Maxwell distribution as a stationary solution. We devote
Appendix~\ref{sec:app_b} to discuss these issues in detail. 

Our paper is organized as follows.
In Sec.~\ref{sec:langevin} we summarize well known results on the diffusion of
non-relativistic heavy quarks in a hot plasma, displaying how the Langevin
equation can be used as an effective theory to deal with the problem. 
However, in high-energy collisions charm and bottom spectra, though rapidly
decreasing, have long power-law tails extending up to very large transverse
momenta. Hence the need of extending the Langevin equation to the relativistic
case. 
In Sec.~\ref{sec:Q} we thus present the calculation of the transport
coefficients for a quark with a generic momentum $p$. In Sec.~\ref{sec:QQbar}
the approach is extended to deal with a $Q\overline{Q}$ pair. 
Sec.~\ref{sec:numres} is devoted to the numerical results both for the
transport coefficients and for the Langevin evolution of heavy-quark
spectra. Finally, in Sec.~\ref{sec:concl} we discuss our results and draw our
conclusions. 
Some technical details are given in the appendices. In Appendix~\ref{app:a} we
recall the essential formulas for the HTL gluon propagators and spectral
functions which are employed in the text. In Appendix~\ref{sec:app_b} we give a
self-contained discussion of the problems one has to face in solving the
relativistic Langevin equation. In Appendix~\ref{sec:app_c} we report the
results of the kinetic calculation of the momentum-diffusion coefficient in the
non-relativistic case, which is used as a benchmark to have some control on the
uncertainties related to our effective HTL approach. 

\section{Langevin approach for heavy quarks in the QGP: a brief summary}
\label{sec:langevin}

The Langevin equation has been recently employed by several authors
\cite{svet,tea,rapp2,rapp3,hira,shury} as an effective theory to describe the
evolution in coordinate and momentum space of heavy quarks in a hot plasma.
In this section we briefly summarize the main aspects of such an approach,
focusing in particular on its range of validity and on the information on
the medium properties that can be extracted from the study of the heavy quark
propagation.

For the sake of simplicity our introductory discussion refers to the non-relativistic case, i.e. with the heavy quarks not too far from thermal equilibrium. This will be sufficient for the above purposes. Actually, in heavy-ion collisions, $c$ and $b$ quarks are produced with a sizable transverse momentum, so that a relativistic study is in order. This however would introduce some technicalities; hence we postpone it to the following sections, where we shall see that, as long as $T/E\ll 1$, the Langevin approach remains justified.

In \cite{hq0,pisar} it was shown that the interaction rate of a very massive
(i.e. with $M\gg T$) quark at rest in a hot plasma is of order $g^2T$ and is
given by (here we consider the QED case):
\beq\label{eq:rate0}
\Gamma=g^2T\int\frac{d\bm{q}}{(2\pi)^3}\frac{\pi m_D^2}{(\bm{q}^2+m_D^2)^2 q}=
\frac{g^2T}{4\pi}.
\eeq
Since the Debye mass $m_D\sim gT$ is the only scale within the integral, the
typical momentum exchanged in the interaction will be of its
order \footnote{Note that the total interaction rate is independent on the value
of $m_D$, which is only relevant to set the typical scale of exchanged momenta
and to prevent infrared divergences by screening the interaction.}.
In particular the maximum of the integrand occurs at
$\overline{q}=m_D/\sqrt{3}$, so that most of the kicks received by the heavy
quark will involve the exchange of soft electrostatic photons (since $M$ is very
large, one can neglect magnetic interactions and consider the static limit) with
momentum, and virtuality, $q\sim gT$. Hence the duration of a collision 
$\tau_\text{coll}\sim 1/gT$ will be well separated, at least in the weak
coupling regime, from the average time interval between two scattering events
$\Delta\tau\equiv 1/\Gamma\sim 1/g^2T$. Since $\tau_\text{coll}\ll\Delta\tau$,
one is then allowed to treat the diffusion of the heavy quarks as resulting from
the sum of many uncorrelated momentum kicks.

Due to the large mass $M$, a huge number of collisions is required for an
ensemble of heavy quarks to thermalize. At thermal equilibrium, from the
equipartition theorem, one has $\overline{p}_{\text{heavy}}=\sqrt{3MT}$,
while for the light particles $\overline{p}_{\text{light}}\sim T$, hence
\beq
\langle p_{\text{heavy}}^2\rangle\sim\frac{M}{T}\langle
p_{\text{light}}^2\rangle, 
\eeq
the same relation expressing the number of random collisions
required to change the average squared-momentum by a factor of order one. 
In the case 
$M\gg T$ the heavy-quark relaxation time will then be much larger than
the one for the light particles of the medium, namely:
\beq\label{eq:est}
\tau_{\text{heavy}}\sim\frac{M}{T}\,\tau_{\text{light}}\underset{g\ll
  1}{\sim}\frac{M}{T}\frac{1}{g^4T\ln(1/g)}, 
\eeq 
where the last estimate
refers to the weak-coupling regime ($\tau_{\text{light}}$ being of the order of
the time interval between two hard collisions) and will be shown to hold in the following. 

In the Langevin approach \cite{langevin} the equation of motion for a
non-relativistic heavy quark reads
\beq\label{eq:lange0}
\frac{dp^i}{dt}=-\eta_D p^i+\xi^i(t),
\eeq
where the right hand side is given by the sum of a friction force (described by
the \emph{drag} coefficient $\eta_D$) and a noise term, which is fixed by its
temporal correlator: 
\beq
\langle\xi^i(t)\xi^j(t')\rangle=\kappa\,\delta^{ij}\delta(t-t').
\eeq
The noise arises from the random uncorrelated momentum kicks received from the
medium. 
Eq.~(\ref{eq:lange0}) can be solved by discretizing the time derivative with a
time-step $\Delta t$ sufficiently large to include many collisions, but much
shorter than the relaxation time of the heavy quark. The latter represents the
typical time-scale over which one would like to follow the momentum
evolution. The condition to be fulfilled is then: 
\beq
  \tau_{\text{light}}\ll\Delta t\ll\tau_{\text{heavy}}\sim\frac{M}{T}\,
    \tau_{\text{light}}.
\eeq 
The \emph{momentum diffusion coefficient} $\kappa$ represents the \emph{averaged
squared momentum} acquired by the heavy quark per unit time,
\beq
\kappa\equiv\frac{1}{3}\langle\frac{\Delta \bm{p}^2}{\Delta t}\rangle,
\eeq
and arises from the cumulated effect of many kicks suffered in the elementary
time-step. 
It can be calculated from Eq.~(\ref{eq:rate0}), by weighting the
differential interaction rate with the squared momentum transfer, thus getting
\beqa
  \kappa&=&\frac{g^2 m_D^2
    T}{6\pi}\int_0^{q_\text{max}}\frac{q^3dq}{(q^2+m_D^2)^2} 
    \nonumber\\
  &=&\frac{g^2 m_D^2 T}{12\pi}\left[\ln\frac{q_\text{max}^2+m_D^2}{m_D^2}-
    \frac{q_\text{max}^2}{q_\text{max}^2+m_D^2}\right].
\eeqa
Notice that, while the interaction rate is free of ultraviolet divergences ---
hard scatterings not occurring very frequently --- in the calculation of
transport coefficients the latter play a major role, since, in spite of being 
quite rare, they can lead to a sizable momentum transfer. 
In the above $q_\text{max}\sim T$ reflects
an estimate of the maximum momentum exchange with a typical thermal
particle. One often considers the result at Leading Logarithmic Accuracy (LLA):
\beq\label{eq:knrlla}
\kappa^{\text{LLA}}=\frac{g^2 m_D^2 T}{6\pi}\int_{m_D}^{q_\text{max}}
\frac{dq}{q}=\frac{g^2 m_D^2 T}{6\pi}\ln\frac{q_\text{max}}{m_D},
\eeq
where the argument of the logarithm is of order $1/g$, its precise numerical
value representing the theoretical uncertainty of the 
present approach.

By requiring the momentum distribution to reach thermal equilibrium one gets
the Einstein relation between the drag and momentum diffusion
coefficients:
\beq\label{eq:einstein_nr}
\eta_D=\frac{\kappa}{2MT}.
\eeq
Concerning the evolution in space, from
\beq
\langle x^2(t)\rangle\underset{t\to\infty}{\sim}6Dt,\quad
\text{with}\quad
x^i(t)=\int_0^tdt'\frac{p^i(t')}{M},
\eeq
one derives the (spatial) \emph{heavy-quark diffusion coefficient}
\beq
D=\frac{T}{M\eta_D}=\frac{2T^2}{\kappa},
\eeq
whose expression at LLA follows from Eq.~(\ref{eq:knrlla})   
\beq\label{eq:DLLA}
D^{\text{LLA}}=\frac{12\pi T}{g^2 m_D^2
\displaystyle{\ln\frac{q_\text{max}}{m_D}}}.
\eeq
For the heavy-quark relaxation time one has then:
\beq\label{eq:hqrel}
\tau_{\text{heavy}}\equiv\frac{1}{\eta_D}=\frac{M}{T}D.
\eeq
By comparing Eqs.~(\ref{eq:est}) and (\ref{eq:hqrel}) one gets the
interesting result \cite{tea}
\beq
D\sim\tau_{\text{light}}\underset{g\ll
  1}{\sim}\frac{1}{g^4T\ln(1/g)},
\eeq
showing that the heavy quark diffusion coefficient gives a reasonable
estimate of the relaxation time of the medium. We also notice that, in the weak
coupling regime, from Eq.~(\ref{eq:DLLA}), one has
$\tau_{\text{light}}\sim 1/g^4T\ln(1/g)$. 

For a recent study of the possibility of estimating heavy-quark transport
coefficients from euclidean lattice simulations see Ref.~\cite{lai}. 

\section{Propagation of a heavy quark in a hot plasma: transport properties}
\label{sec:Q}

We start our investigation by considering the propagation of a heavy quark in
a hot ultra-relativistic plasma. When the quark is very massive and/or is 
endowed with a very large momentum $p\gg T$, it is reasonable to
describe it within the eikonal approximation, in which it moves along a
straight-line trajectory, acquiring a phase due to the interaction with the
background gauge field.

One could argue that, due to the multiple kicks received from the particles
of the medium,
during its propagation the heavy quark would loose energy and acquire more and
more transverse momentum, so that at some point the assumption of
straight-line propagation should cease to be valid. However,
one can take advantage of the huge separation, occurring for $E_p\gg T$,
between the relaxation time of the medium,  $\tau_{\text{light}}$, and the much
larger time required by the heavy quark to approach equilibrium.
In order to evaluate transport coefficients (related for instance to 
heavy-quark energy loss, diffusion and momentum broadening),
it is sufficient to follow the propagation of the quark for a time large
compared to $\tau_{\text{light}}$ but still much shorter than the relaxation
time of the high-momentum heavy quark. For this purpose 
our eikonal approach should be justified and indeed turns out to provide
results in agreement (at least at Leading Logarithmic 
Accuracy) with the ones obtained in other approaches as, for example, by
solving the Maxwell equations in a dielectric medium \cite{Mat2,gyu,mus}.
If one is really interested in following the relaxation of the heavy quarks
toward thermal equilibrium, then one can use the above findings
for the transport coefficients and insert them into the Langevin or
Fokker-Planck equations. 

Following the approach developed in Refs.\cite{hq0,hq1}, the configurations of
the gauge-field entering into the eikonal phase are weighted by the 
HTL effective action. We start from the QED case (i.e. a plasma
of photons, electrons and positrons), in which the latter is gaussian; this
allows to perform the functional integral exactly, leading to an exact
exponentiation of the gauge-field propagator. Later we will show
how the results for the transport properties
can be generalized to the QCD case, yet recovering well known findings.

In analogy with the approach adopted in Refs.~\cite{hq0,BNbla},
we consider the retarded propagator of a heavy quark
--- which is treated as a test particle --- created at
$(0,\bm{r}_1')$ and annihilated at $(t,\bm{r}_1)$; it is defined as
\beq
G^R(t,\bm{r}_1|0,\bm{r}_1')\equiv i\,\theta(t)\langle\psi(t,\bm{r}_1)
\psi^\dagger(0,\bm{r}_1')\rangle,
\eeq
where the expectation value refers to a thermal average over the states of a
hot medium of light particles. In the eikonal approximation it is given by:
\beq
  G^R(t,\bm{r}_1|0,\bm{r}_1')=i\,\theta(t)\,
    \delta(\bm{r}_1-\bm{r}_1'-\bm{v} t)\overline{G}(t).
\eeq
In the above, the eikonal phase (which is complex, the imaginary part accounting
for the effects of collisions) is expressed in terms of
the \emph{real-time} gauge-field propagator given in Appendix~\ref{app:a}:
\beq\label{eq:eikphase}
\overline{G}(t)=
\exp\left[\,\frac{i}{2}\int d^4x \int d^4y\,J^\mu(x) D_{\mu\nu}(x-y)
J^\nu(y)\right],
\eeq
where the current describing the propagation of the heavy quark is given by:
\beq
\label{eq:Qcurrent}
  J^\mu(x)=
    g\theta(x^0)\theta(t-x^0)\delta(\bm{x}-\bm{r}_1'-\bm{v} x^0)(1,\bm{v}).
\eeq
After expressing the gauge propagator in Fourier space one gets:
\begin{multline}
\overline{G}(t)=
\exp\Big\{\,\frac{i}{2}g^2\int\frac{d\omega}{2\pi}\int\frac{d\bm{q}}{(2\pi)^3}
\frac{2\left[1-\cos(\omega-\bm{q}\cdot\bm{v})t\right]}
{(\omega-\bm{q}\cdot\bm{v})^2} \\ 
\times \left[D_L(\omega,q)+v^2\left(1-(\hat{\bm{v}}
\cdot\hat{\bm{q}})^2\right)D_T(\omega,q)\right] \Big\},
\end{multline}
whose large-time behavior can be obtained from the limit
\beq
\lim_{t\to\infty}\frac{1-\cos(\omega-\bm{q}\cdot\bm{v})t}
{(\omega-\bm{q}\cdot\bm{v})^2}=\pi t \delta(\omega-\bm{q}\cdot\bm{v}).
\label{eq:limcos}
\eeq
In particular the probability of finding the heavy quark with
momentum $\bm{p}=\gamma M\bm{v}$ will tend to decrease due to the collisions
with the plasma particles, whose effects are encoded into the imaginary part of
the HTL gauge-field propagator (see Appendix~\ref{app:a}):
\beq
 \text{Im}\,D_{L/T}(\omega,q)=\rho_{L/T}(\omega,q)
\left(N(\omega)+\frac{1}{2}\right).
\eeq
This allows to \emph{formally} define an interaction rate for a quark
propagating with velocity $\bm{v}$ (taken in the following along the $z-$axis),
which turns out to be given by (the spectral functions being odd)
\beqa
  \Gamma&=&g^2\int d\omega\int\frac{d\bm{q}}{(2\pi)^3}
  \delta(\omega-\bm{q}\cdot\bm{v}) \left[\rho_L(\omega,q)+
    v^2\left(1-(\hat{\bm{v}}\cdot\hat{\bm{q}})^2\right)\rho_T(\omega,q)\right]
    N(\omega)\nonumber\\
  {}&\equiv&g^2\int d\omega\int\frac{d\bm{q}}{(2\pi)^3}
    \delta(\omega-\bm{q}\cdot\bm{v})\widetilde{\rho}(\omega,\bm{q})N(\omega).
\label{eq:rate}
\eeqa
Note that, strictly speaking, the above expression develops an infrared
divergence arising from the exchange of long-wavelength magneto-static
gluons\footnote{Actually, lattice QCD results provide evidence for the existence of a non vanishing \emph{magnetic mass}~\cite{magn}, of non-perturbative origin, which would eliminate the above divergence. This in principle could affect our results for the transport coefficients, but such an issue lies beyond the scopes of the present paper.}. However the 
above processes cannot affect the transport properties related to the
heavy-quark propagation (energy loss, momentum broadening, ...), since they are
related to negligible energy/momentum exchanges. 
Hence we can take advantage of Eq.~(\ref{eq:rate}) for the computation of
the transport coefficients we are interested in.

The energy loss per unit length is then given by:
\beq
\frac{dE}{dx}
=\frac{g^2}{v}
\int d\omega\int\frac{d\bm{q}}{(2\pi)^3}
\delta(\omega-\bm{q}\cdot\bm{v})
\widetilde{\rho}(\omega,\bm{q})\,
\omega
N(\omega).
\eeq
The Dirac delta
\beq
\delta(\omega-\bm{q}\cdot\bm{v})=\frac{1}{qv}\,\delta\left(\cos\theta-
\frac{\omega}{qv}\right)
\eeq
can be exploited to perform the angular integration.
Furthermore, since the spectral function is odd, one can replace the Bose
distribution with its even part
\beq
N(\omega)\rightarrow\frac{N(\omega)+N(-\omega)}{2}=-\frac{1}{2},
\eeq
thus obtaining
\beqa
  -\frac{dE}{dx}&=&\frac{g^2}{4\pi^2v^2}\int_0^{q_\text{max}}dq\,q\int_{0}^{vq}
    d\omega\,\omega  
    \left[\rho_L(\omega,q) + \left(v^2-\frac{\omega^2}{q^2}\right)
    \rho_T(\omega,q)\right] \nonumber\\
  {}&\equiv&\frac{g^2}{4\pi^2v^2}\int_0^{q_\text{max}}q\,dq\int_{0}^{vq}
  d\omega\,\omega\overline{\rho}(\omega,q),\label{eq:elossQ}
\eeqa
where, in the high-energy limit, $q_\text{max}\sim\sqrt{ET}$, with $E$ the
energy of the heavy quark, is the maximum momentum transfer in a collision with
a plasma particle, whose typical momentum is of order $T$. At LLA the energy
loss can be computed analytically, by setting a lower bound of order $m_D$ in
the momentum  integration and using for the spectral functions the expressions
given in  Eq.~(\ref{eq:LLAsp}). One gets:
{\setlength\arraycolsep{1pt}
\beqa
  \left.-\frac{dE}{dx}\right|_\text{LLA}&=&\frac{g^2m_D^2}{4\pi v^2}
    \int_{m_D}^{q_\text{max}}\frac{dq}{q^4}
    \int_{0}^{vq}d\omega\,\omega^2 \left[1+\frac{v^2-\omega^2/q^2}{2
    \left(1-\omega^2/q^2\right)}\right] \nonumber\\
  {}&=&\frac{g^2m_D^2}{4\pi v^2}
    \int_{m_D}^{q_\text{max}}\frac{dq}{q}\int_{0}^{v}dx\,x^2
    \left[1+\frac{v^2-x^2} {2(1-x^2)}\right]. \nonumber\\
\eeqa}
After performing the integrals one gets:
\beq\label{eq:eLLA}
\left.-\frac{dE}{dx}\right|_\text{LLA}=\frac{g^2m_D^2}{8\pi v}
\ln\frac{q_\text{max}}{m_D}\left[1-\frac{1-v^2}{2v}\ln\frac{1+v}{1-v}\right],
\eeq
to be compared with the analogous result displayed in Eq.~(B31) of 
Ref.~\cite{tea} for the QCD case and quoting previous findings obtained in 
Refs.~\cite{tho1,tho2}. 

We now address the calculation of the momentum diffusion coefficients $\kappa_T$
and $\kappa_L$ which will enter the noise term in the Langevin equation [see 
Eqs.~(\ref{eq:noise1}) and (\ref{eq:noise2})].

Let us start with the transverse momentum diffusion coefficient,
representing the mean squared transverse momentum acquired per unit time
by the heavy quark crossing the medium, namely:
\beq
\kappa_T\equiv\frac{1}{2}\langle\frac{\Delta p_T^2}{\Delta t}\rangle.
\eeq
Also the latter can be computed starting from Eq.~(\ref{eq:rate}):
\beq
  \kappa_T=\frac{g^2}{2}\int d\omega\int\frac{d\bm{q}}{(2\pi)^3}
    \delta(\omega-\bm{q}\cdot\bm{v})\widetilde{\rho}(\omega,\bm{q}) 
     q^2(1-\cos^2\theta)N(\omega).
\eeq
In the above only the odd part of the Bose distribution contributes to the
integral, so that one can make the replacement
\beq
N(\omega)\rightarrow\frac{N(\omega)-N(-\omega)}{2}=\frac{1}{2}
\coth\frac{\beta\omega}{2},
\eeq 
thus obtaining:
\beq
  \kappa_T=\frac{g^2}{8\pi^2v}\int_0^{q_\text{max}}dq\,q\int_{0}^{vq}
    d\omega\,\overline{\rho}(\omega,q)
    q^2\left(1-\frac{\omega^2}{q^2v^2}\right) \coth\frac{\beta\omega}{2}.
\eeq
In analogy with the energy loss calculation, the LLA result for the transverse
momentum diffusion is obtained by employing the
corresponding approximate expression for the HTL spectral functions given
in Appendix~\ref{app:a}, setting an infrared cutoff of order $m_D$ in the
momentum integration and assuming that the relevant contribution to
the above expression is given by processes with small energy transfer, so that
one can approximate
\beq
\coth\frac{\beta\omega}{2}\sim\frac{2T}{\omega}.
\eeq
One obtains:
{\setlength\arraycolsep{1pt}
\beqa
\kappa_T^{\text{LLA}}&=&\frac{g^2m_D^2}{8\pi v}
\int_{m_D}^{q_\text{max}}\frac{dq}{q^2}\int_{0}^{vq}d\omega\,\omega
\left[1+\frac{v^2-\omega^2/q^2}{2\left(1-\omega^2/q^2\right)}\right]
\left(1-\frac{\omega^2}{q^2v^2}\right)\frac{2T}{\omega}
\nonumber\\
{}&=&\frac{g^2Tm_D^2}{4\pi v}
\int_{m_D}^{q_\text{max}}\frac{dq}{q}\int_{0}^{v}dx\,\left[1+\frac{v^2-x^2}
{2(1-x^2)}\right]\left(1-\frac{x^2}{v^2}\right) .
\eeqa}
After performing the integrals one gets
\beq
\kappa_T^{\text{LLA}}\!=\!\frac{g^2Tm_D^2}{4\pi}
\ln\frac{q_\text{max}}{m_D}\left[\frac{3}{2}-\frac{1}{2v^2}+
\frac{(1-v^2)^2}{4v^3}\ln\frac{1+v}{1-v}\right],
\eeq
again in agreement with the leading log term in Eq.~(B32) of Ref.~\cite{tea}.
Note that the transverse momentum diffusion is closely related to the transport
coefficient $\hat{q}$, often entering into the radiative energy loss
calculation as a parameter encoding the properties of the medium, and
representing the \emph{transverse squared momentum} acquired by the propagating
parton \emph{per mean free path} \cite{baier}. 
One can in fact measure the distance $X$ covered in the medium in units of
$\lambda_\text{mfp}$, setting $X\equiv n\lambda_\text{mfp}$,
so that
\beq
\Delta X\equiv \lambda_\text{mfp}.
\eeq
Hence
\beq
\hat{q}\equiv\frac{\langle\Delta p_T^2\rangle}{\lambda_\text{mfp}}
=\langle\frac{\Delta p_T^2}{\Delta X}\rangle=
\frac{1}{v}\langle\frac{\Delta p_z^2}{\Delta t}\rangle
=\frac{2}{v}\kappa_T,
\eeq   
leading, in the case of an ultra-relativistic parton, to
\beq
\hat{q}=\frac{g^2Tm_D^2}{2\pi}
\ln\frac{q_\text{max}}{m_D},
\eeq
in agreement with the findings of Ref.~\cite{yac}.

The calculation of the longitudinal momentum diffusion
\beq
\kappa_L\equiv\langle\frac{\Delta p_z^2}{\Delta t}\rangle
\eeq
follows in perfect analogy. From
\beq
\kappa_L=g^2
\int d\omega\int\frac{d\bm{q}}{(2\pi)^3}\delta(\omega-\bm{q}\cdot\bm{v})
\widetilde{\rho}(\omega,\bm{q})\,
q^2\cos^2\theta\,N(\omega)
\eeq
one gets
\beq
\kappa_L=\frac{g^2}{4\pi^2v}\int_0^{q_\text{max}}q\,dq\int_{0}^{vq}
d\omega\,\overline{\rho}(\omega,q)
\frac{\omega^2}{v^2}
\coth\frac{\beta\omega}{2}.
\eeq
Again the result at LLA can be expressed analytically:
{\setlength\arraycolsep{1pt}
\beqa
  \kappa_L^{\text{LLA}}&=&\frac{g^2Tm_D^2}{2\pi v^3}
    \int_{m_D}^{q_\text{max}}\frac{dq}{q^4}\int_{0}^{vq}d\omega\,\omega^2 
     \left[1+\frac{v^2-\omega^2/q^2}
    {2\left(1-\omega^2/q^2\right)}\right] \\
  {}&=&\frac{g^2Tm_D^2}{2\pi v^3}
    \int_{m_D}^{q_\text{max}}\frac{dq}{q}\int_{0}^{v}dx\,x^2
    \left[1+\frac{v^2-x^2}{2(1-x^2)}\right], \nonumber
\eeqa}
which leads to:
\beq\label{eq:klLLA}
\kappa_L^{\text{LLA}}=\frac{g^2Tm_D^2}{4\pi v^2}
\ln\frac{q_\text{max}}{m_D}\left[1-\frac{1-v^2}{2v}\ln\frac{1+v}{1-v}\right],
\eeq
to be compared with Eq.~(B33) of Ref.~\cite{tea}.
Note that at LLA the relation between friction and momentum-diffusion
coefficients, required by the Langevin approach (see Sec.~\ref{sec:langevin}
and App.~\ref{sec:app_b}), holds: 
\beq
\kappa_L^{\text{LLA}}=\frac{2T}{v}\left|\frac{dp}{dt}\right|=
\frac{2T}{v}\left|\frac{dE}{dx}\right|_\text{LLA}=2TE\eta_D^{\text{LLA}},
\eeq
as it can be seen by comparing Eqs.~(\ref{eq:eLLA}) and (\ref{eq:klLLA}).
We notice however that, as it will be discussed in the following, when the
strength of the noise depends on the momentum of the brownian particle (as it
turns out to be the case) the viscous term in the Langevin description needs a
correction (subleading in T/E) in order to recover the correct continuum limit
to the Fokker-Planck equation. 

Let us now check \emph{a posteriori} that in a time interval of order
$\tau_{\text{light}}$ the momentum loss $\Delta p$ and the acquired
$\sqrt{\langle p_T^2\rangle}$
are negligible with respect to the initial $E_p\gg T$, so that the calculation
of transport coefficients within the eikonal approach results meaningful.

From $dp/dt=dE/dx$, during the time $\tau_{\text{light}}$, one gets:
\beq
|\Delta p|=
\left|\frac{dE}{dx}\right|\tau_{\text{light}}
\sim\left|\frac{dE}{dx}\right|\,D
\sim\frac{g^2m_D^2}{8\pi}
\ln\frac{q_\text{max}}{m_D}\,\frac{12\pi T}{g^2 m_D^2
\displaystyle{\ln\frac{q_\text{max}}{m_D}}}\sim T \ll p.
\eeq
Concerning the transverse momentum, during the time
interval $\tau_{\text{light}}$, one has  
\beq
\langle \Delta p_T^2\rangle=2\kappa_T\tau_{\text{light}}\sim\kappa_T D,
\eeq
so that
\beqa
  \sqrt{\langle \Delta p_T^2\rangle}&\sim&\left(
    \frac{g^2Tm_D^2}{4\pi}\ln\frac{q_\text{max}}{m_D}\frac{12\pi T}{g^2 m_D^2
    \displaystyle{\ln\frac{q_\text{max}}{m_D}}}\right)^{1/2} \nonumber\\
  &\sim& T\ll p.
\eeqa

\section{Transport properties for a $Q\overline{Q}$ pair}\label{sec:QQbar}

Within the same framework it is possible to study the transport properties of a
quarkonium, which we simply model as a $Q\overline{Q}$ pair, of a fixed size,
propagating with velocity $\bm{v}$ in a hot plasma.
Similar studies, limited to the energy-loss problem and to the definition of a
``dipole potential'', can be found in Refs.~\cite{Mat2,mus}.
Here the problem is addressed by solving the Maxwell equations in the linear response approximation for a $Q\overline{Q}$ pair (playing the role of an external current) propagating in the QGP, modeled as a dielectric medium.
 
Clearly, the results will depend not only on the size of the ``dipole'',
but also on its orientation with respect to the direction of propagation.
Following the choice adopted in Ref.~\cite{mus}, we take, without loss of
generality, $\bm{v}$ along the $z$-axis, $\bm{r}\!\equiv\!(r_\perp,0,z)$ in the 
$zx$-plane and the exchanged momentum as
$\bm{q}\!=\!q(\sin\theta\cos\phi,\sin\theta\sin\phi,\cos\theta)$.

We consider then the creation of a quark at $(0,\bm{r}_1')$ and of an anti-quark
at $(0,\bm{r}_2')$, with $\bm{r}\equiv\bm{r}_1'-\bm{r}_2'$, and we
follow the propagation of this pair within the eikonal approximation, as
already done for the single particle case in Sec.~\ref{sec:Q}.
Hence, one can take advantage of Eq.~(\ref{eq:eikphase}), where now
the current describes the propagation of a dipole and is given by,
\beq
  J^\mu(x)=
    g\theta(x^0)\theta(t-x^0) 
  \left[\delta(\bm{x}-\bm{r}_1'-\bm{v} x^0) 
    -\delta(\bm{x}-\bm{r}_2'-\bm{v} x^0)\right](1,\bm{v}),
\eeq
so that for the $Q\overline{Q}$ propagator one gets:
\begin{eqnarray}
  \overline{G}(t)&=&
    \exp\Big\{ig^2\int\frac{d\omega}{2\pi}\int\frac{d\bm{q}}{(2\pi)^3}
    \frac{2[(1-\cos(\omega-\bm{q}\cdot\bm{v})t]}
    {(\omega-\bm{q}\cdot\bm{v})^2} \nonumber\\
  &&\times\left[1-\cos(\bm{q}\cdot\bm{r})\right]
    \left[D_L(\omega,q)+v^2\left(1-(\hat{\bm{v}}
\cdot\hat{\bm{q}})^2\right)D_T(\omega,q)\right]
\Big\}. 
\end{eqnarray}
One can again consider the large-time decay of the above correlator, which
allows to identify the interaction rate of the dipole: 
\beq
  \Gamma=2g^2\int d\omega\int
    \frac{d\bm{q}}{(2\pi)^3}\left[1-\cos(\bm{q}\cdot\bm{r})\right]
    \delta(\omega-\bm{q}\cdot\bm{v}) \widetilde{\rho}(\omega,\bm{q})N(\omega).
\label{eq:rate_dip1}
\eeq
The Dirac delta can be used to perform the integral over the polar angle,
while the azimuthal integration gives rise to a Bessel function,
by exploiting the integral representation
\beq
J_0(q_\perp r_\perp)=\int_0^{2\pi}\frac{d\phi}{2\pi}
e^{\displaystyle{iq_\perp r_\perp\cos\phi}},
\eeq
with $q_\perp\equiv q\sin\theta$.
One can then study the transport properties for the $Q\overline{Q}$ pair.

We first examine the energy loss which is given by [see Eq.~(\ref{eq:elossQ})]:
\beq
\label{eq:elosspair}
  -\frac{dE}{dx} = \frac{g^2}{2\pi^2 v^2}\int_0^{q_\text{max}} q\,dq\int_{0}^{qv}
    d\omega \left[1-J_0\left(q r_\perp\sqrt{1-\frac{\omega^2}{q^2v^2}}\right)
\cos\left(\frac{z\omega}{v}\right)\right]\overline{\rho}(\omega,q)\,\omega.
\eeq
For the transverse and longitudinal momentum diffusion coefficients one gets
\beqa
  \kappa_T &=& \frac{g^2}{4\pi^2 v}\int_0^{q_\text{max}} q\,dq\int_{0}^{qv}
    d\omega \left[1-J_0\left(q r_\perp\sqrt{1-\frac{\omega^2}{q^2v^2}}\right)
    \cos\left(\frac{z\omega}{v}\right)\right] \nonumber \\ 
  &&\quad\times\overline{\rho}(\omega,q)\,\coth\left(\frac{\beta\omega}{2}\right)
q^2\left(1-\frac{\omega^2}{q^2v^2}\right)
\eeqa
and
\beqa
  \kappa_L &=& \frac{g^2}{2\pi^2 v}\int_0^{q_\text{max}} q\,dq\int_{0}^{qv}
    d\omega\left[1-J_0\left(q r_\perp\sqrt{1-\frac{\omega^2}{q^2v^2}}\right)
    \cos\left(\frac{z\omega}{v}\right)\right] \nonumber \\
  && \quad\times \overline{\rho}(\omega,q)\,
\coth\left(\frac{\beta\omega}{2}\right)
\frac{\omega^2}{v^2},
\eeqa
respectively.

In the calculation one also needs to account for the dependence on the dipole
orientation. For this purpose we start by considering 
the \emph{rest frame of the propagating pair} (whose coordinates we label with
a tilde): for a $S-$wave state (till very large velocities the in-medium
potential is still approximately spherically symmetric \cite{Mat3}) one has 
\beq
R^2=\langle \tilde{x}^2\rangle+\langle \tilde{y}^2\rangle+\langle
\tilde{z}^2\rangle. 
\nonumber
\eeq
Moving then to the \emph{rest frame of the thermal bath} (the laboratory frame),
if the pair propagates along the $z-$axis, its transverse size (in the
$xy-$plane) is not affected and one has:
\beq
r_\perp\equiv\sqrt{\langle x^2\rangle+\langle y^2\rangle}=
\sqrt{\langle \tilde{x}^2\rangle+\langle \tilde{y}^2\rangle}
=\sqrt{2/3}R.
\nonumber
\eeq  
On the other hand, the longitudinal size of the dipole turns out to be
Lorentz-contracted, according to the relation
\beq
z=\tilde{z}/\gamma=R/(\sqrt{3}\gamma).
\eeq
The above expressions of $r_\perp$ and $z$ will be employed in the calculation
of the transport coefficients. 

The small dipole limit can be easily estimated by considering the
expansion 
\beqa
1-J_0(q r_\perp\sin\theta)\cos(qz\cos\theta)&
\underset{R\ll q_\text{max}^{-1}}{\sim}
&\frac{q^2 r_\perp^2\sin^2\theta}{4}+\frac{q^2z^2\cos^2\theta}{2}\nonumber\\
{}&=&\frac{q^2R^2}{6}\left(\sin^2\theta+\frac{\cos^2\theta}{\gamma^2}\right);
\eeqa
hence, for small sizes, the transport coefficients grow linearly with the
transverse area of the dipole. 
One gets then, in the limits of very small and very large velocities,
\beqa
\left\langle-\frac{dE}{dx}\right\rangle_{v\ll 1}&
\underset{R\ll q_\text{max}^{-1}}
{\sim}&
\frac{g^2R^2}{12\pi^2 v^2}\int_0^{q_\text{max}} q^3\,dq
\int_{0}^{qv}
d\omega \,\overline{\rho}(\omega,q)\, \omega\nonumber\\
\left\langle-\frac{dE}{dx}\right\rangle_{v\to 1}&
\underset{R\ll q_\text{max}^{-1}}
{\sim}&
\frac{g^2R^2}{12\pi^2}\int_0^{q_\text{max}} q^3\,dq
\int_{0}^{q}
d\omega \,\overline{\rho}(\omega,q)\, 
\left(1-\frac{\omega^2}{q^2}\right)\omega
\eeqa
for the energy loss,
\beqa
\left\langle\kappa_T\right\rangle_{v\ll 1}&
\underset{R\ll q_\text{max}^{-1}}{\sim}&
\frac{g^2R^2}{24\pi^2 v}\int_0^{q_\text{max}} q^3\,dq
\int_{0}^{qv}
d\omega \,\overline{\rho}(\omega,q)\,\coth\left(\frac{\beta\omega}{2}\right)
q^2\left(1-\frac{\omega^2}{q^2v^2}\right) \nonumber \\
\left\langle\kappa_T\right\rangle_{v\to 1}&
\underset{R\ll q_\text{max}^{-1}}{\sim}&
\frac{g^2R^2}{24\pi^2}\int_0^{q_\text{max}} q^3\,dq
\int_{0}^{q}
d\omega \,\overline{\rho}(\omega,q)\,\coth\left(\frac{\beta\omega}{2}\right)
q^2\left(1-\frac{\omega^2}{q^2}\right)^2 
\eeqa
and
\beqa
\left\langle\kappa_L\right\rangle_{v\ll 1}&
\underset{R\ll q_\text{max}^{-1}}{\sim}&
\frac{g^2R^2}{12\pi^2 v}\int_0^{q_\text{max}} q^3\,dq
\int_{0}^{qv}
d\omega \,\overline{\rho}(\omega,q)\,\coth\left(\frac{\beta\omega}{2}\right)
\frac{\omega^2}{v^2} \\
\left\langle\kappa_L\right\rangle_{v\to 1}&
\underset{R\ll q_\text{max}^{-1}}{\sim}&
\frac{g^2R^2}{12\pi^2 v}\int_0^{q_\text{max}} q^3\,dq
\int_{0}^{q}
d\omega \,\overline{\rho}(\omega,q)\,\coth\left(\frac{\beta\omega}{2}\right)
\omega^2\left(1-\frac{\omega^2}{q^2}\right), \nonumber
\eeqa
for the transverse and longitudinal momentum diffusion coefficients,
respectively. Notice that for $v\to1$ terms containing $z$ are neglected, since 
they are suppressed by the Lorenz $\gamma$ factor: in the ultra-relativistic
limit only the transverse area is relevant for the transport properties. 
In practice, however, the above expansions are not so useful, since the results
depend quadratically on the momentum cutoff $q_\text{max}$, which is not well
determined and gives a huge systematic theoretical uncertainty. On the contrary
working with the exact formulas provides transport coefficients with only a
mild logarithmic dependence on the ultraviolet cutoff. 

\section{Numerical results}\label{sec:numres}

Here we present our numerical results, starting from the transport coefficients
$\kappa_T(p)$ and $\kappa_L(p)$ introduced in Secs.~\ref{sec:Q} and 
\ref{sec:QQbar}; they are then employed in the Langevin evolution of the
momenta of a large sample of heavy quarks (charm and bottom). 
The rigorous results obtained in Sec.~\ref{sec:Q} are generalized to the QCD
case by adding a color charge to the current given in Eq.~(\ref{eq:Qcurrent}),
setting 
\beq
  J^{\mu a}(x)=
    q^a g\theta(x^0)\theta(t-x^0)\delta(\bm{x}-\bm{r}_1'-\bm{v} x^0)(1,\bm{v}),
\eeq
with $a=1,\dots,N_c^2-1$; the corresponding transport
coefficients can be simply obtained by multiplying the ones derived in
Sec.~\ref{sec:Q} by the Casimir factor $C_F\equiv q^aq^a$. 
In the present exploratory study the QGP is modeled as a uniform static medium,
considered at different temperatures spanning a range of interest for heavy ion
collision experiments. 

As already mentioned, the results obtained within the present framework display
a logarithmic dependence on the ultraviolet cutoff $q_\text{max}$, related to
the maximum momentum exchanged in a collision with a typical thermal particle. 
Actually, this represents an intrinsic uncertainty of our scheme,
which --- in dealing with large momentum transfer processes --- should be
supplemented by a microscopic kinetic calculation involving the evaluation of the Born matrix elements for the scattering of the heavy quarks with the particles of the medium \cite{tea,tho1,tho2}.  
However, for the sake of ease and consistency we prefer to pursue the approach
developed in Sec.~\ref{sec:Q}: indeed, it is able to catch all the 
essential qualitative features of the transport coefficients
(e.g. the growth of $\kappa_{T/L}$ with the momentum $p$, its suppression in
the quarkonium case, etc.) and is simple enough (also in view of a more
realistic study of an expanding medium) to be implemented into a Langevin
simulation. 
Bearing in mind that the ambiguity related to the choice of 
$q_\text{max}$ is unavoidable, we refer the reader to Appendix~\ref{sec:app_c}
for details on the procedure employed by us to fix a realistic value for it. 

Concerning the value of $\alpha_s$, which one expects to be quite large at RHIC
and to approach a weak-coupling regime at the (larger) LHC temperatures, it has
been taken from the 2-loop QCD $\beta$-function with the parameters given in
Ref.~\cite{za}. One has 
\beq
g^{-2}(\mu)=2\,b_0\ln\left(\mu/\Lambda_\text{QCD}\right)+
\frac{b_1}{b_0}\ln\big[2\ln(\mu/\Lambda_\text{QCD})\big],
\eeq
with
\beq
b_0=\frac{1}{16\pi^2}\left(11-\frac{2}{3}N_f\right),\;\;
b_1=\frac{1}{(16\pi^2)^2}\left(102-\frac{38}{3}N_f\right)
\eeq
and $\Lambda_\text{QCD}=261$~MeV.
The QCD coupling was then evaluated at a scale proportional to the
temperature. As two representative cases we present the results corresponding
to the typical choices $\mu=\pi T$ and $\mu=2\pi T$, the differences reflecting
the lack of predictivity of the calculation. 
A more careful study of the running coupling effects (actually in the
evaluation of collisional energy-loss) was given in Refs.~\cite{pei1,pei2}, but 
this goes beyond the scope of the present analysis. 

\begin{figure}
\begin{center}
\includegraphics[clip,width=0.5\textwidth]{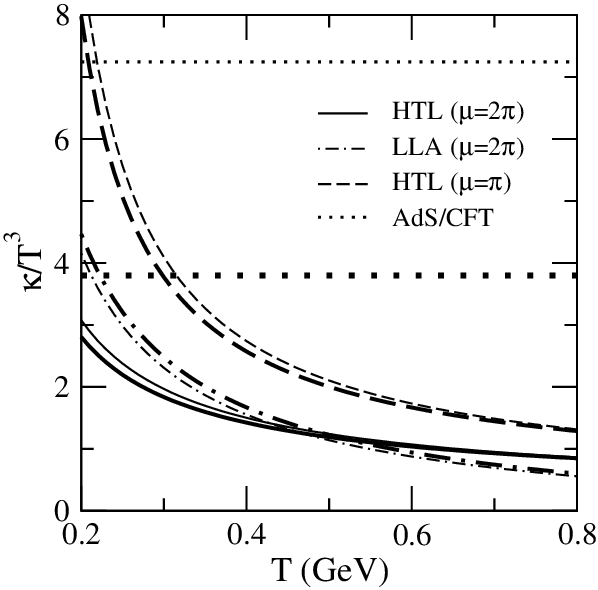}
\caption{The HTL momentum-diffusion coefficients $\kappa_{T/L}$ divided by
  $T^3$ for a bottom quark of mass $M=4.2$ GeV and momentum $p=4$~GeV as a
  function of the temperature. The HTL results, for $\mu=2\pi$ and $\mu=\pi$,
  are shown (solid and dashed lines, respectively). The dot-dashed lines
  correspond to the LLA results. Thick lines refer to $\kappa_T$, thin lines to
  $\kappa_L$. The AdS/CFT values are also reported.}  
\label{fig:0}
\end{center}
\end{figure}

We also compare our results, both for the transport coefficients and for the
Langevin dynamics of the heavy quarks, with the findings provided by the
AdS/CFT correspondence, which, for the $\mathcal{N}=4$ SYM theory, predicts:  
\cite{sol2,gub,raja}
\beqa
\kappa_T&=&\gamma^{1/2}\sqrt{\lambda}\pi T_\text{SYM}^3,\nonumber\\
\kappa_L&=&\gamma^{5/2}\sqrt{\lambda}\pi T_\text{SYM}^3,\label{eq:AdS_kk}
\eeqa
where $\gamma=1/\sqrt{1-v^2}$ and $\lambda=g_\text{SYM}^2N_c$.
Actually, translating the above results to the hot-QCD case is not free of
ambiguities and different strategies were proposed. Here we follow the recipe
adopted in Ref.~\cite{gub}, which is based on matching the energy density and
the $Q\overline{Q}$ force in the two theories, namely: 
\beq
3^{1/4}T_\text{SYM}=T_{QCD}\quad \text{and}\quad\lambda=5.5,
\eeq
the latter being chosen within the range $3.5<\lambda<8$ allowed by the above
mentioned procedure. 
It was pointed out in Ref.~\cite{sol2} that the results in
Eq.~(\ref{eq:AdS_kk}) only hold 
for heavy-quark momenta corresponding to $\gamma$ factors below a critical
limit:
\beq\label{eq:cond}
\gamma<\gamma_c\equiv\left(\frac{M}{\sqrt{\lambda}T_\text{SYM}}\right)^2.
\eeq
To our knowledge no calculation is available for larger momenta.
Hence, in our plots, as $\gamma$ exceeds this critical value, we freeze
$\kappa_T$ and $\kappa_L$ to their estimates below $\gamma_c$.
Had we trivially identified the coupling and the temperature in
the two theories, requiring $\lambda=6\pi$ to reproduce a QCD plasma with
$N_c=3$ and  $\alpha_s=g^2/4\pi\approx 0.5$, the range of validity
of the AdS/CFT calculation would have been even more limited. 

In the range of temperatures covered by our analysis the transport coefficients
$\kappa_{T/L}$ change dramatically, increasing approximately as $T^3$
(deviations being due to running-coupling effects), as it is displayed in
Fig.~\ref{fig:0}. This leads to observable consequences for the heavy-quark
dynamics, that we will discuss. Fig.~\ref{fig:0} refers to the case of a
$b$-quark of not too large momentum, so that the condition in
Eq.~(\ref{eq:cond}) is satisfied. In the same figure we also show for
comparison the LLA result for $\mu=2\pi$.

\begin{figure}
\begin{center}
\includegraphics[clip,width=0.65\textwidth]{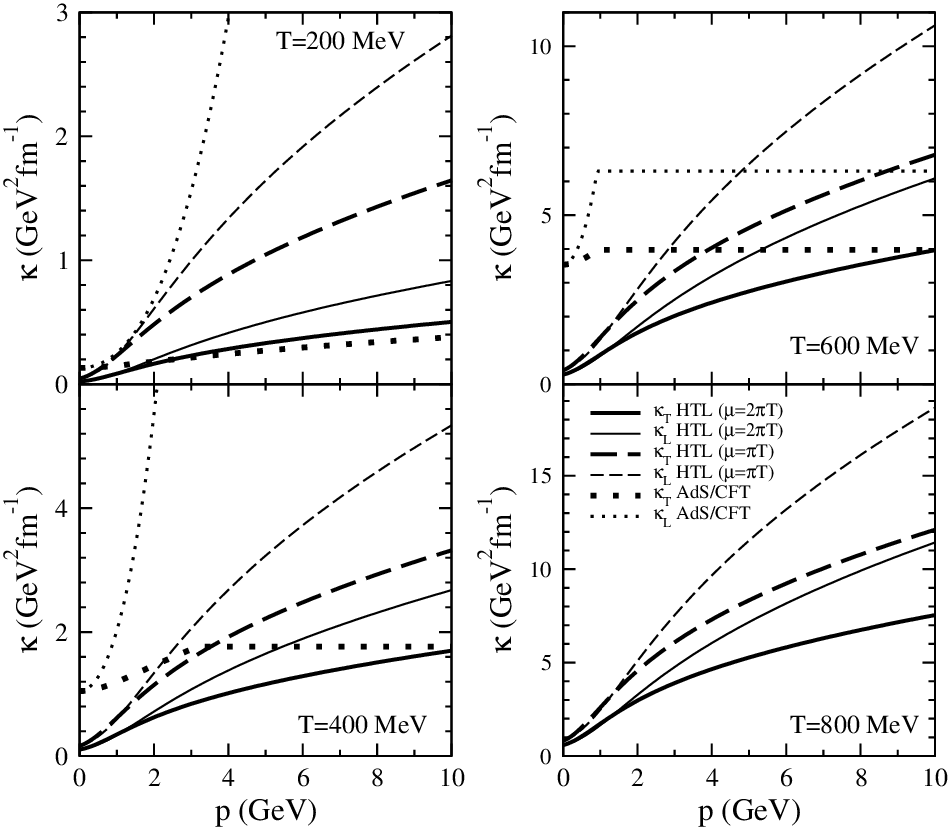}
\caption{The HTL momentum-diffusion coefficients $\kappa_{T/L}(p)$ for a charm
  quark of mass $M=1.2$ GeV for a range of temperatures above $T_c$ of
  experimental interest. We also plot, in the regime where they are available,
  the AdS/CFT results. Again, thick lines refer to $\kappa_T$ and thin ones to
  $\kappa_L$.}  
\label{fig:1}
\vskip 0.3cm
\includegraphics[clip,width=0.65\textwidth]{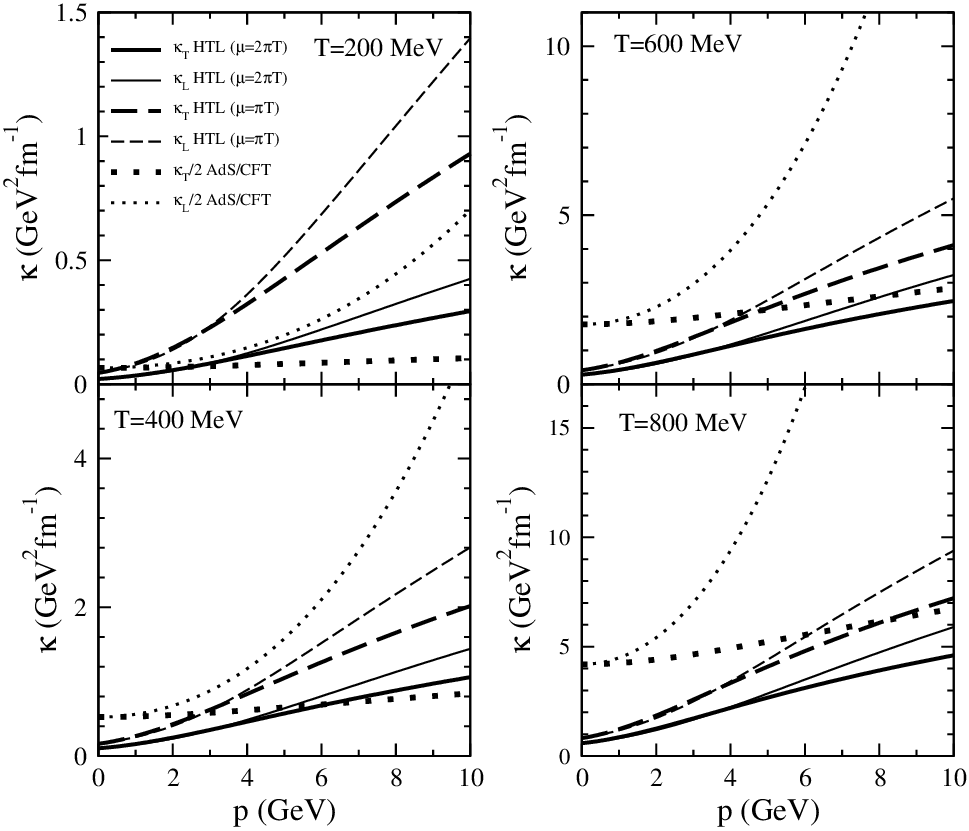}
\caption{As in Fig.~\ref{fig:1}, but for a bottom quark of mass $M=4.2$ GeV.} 
\label{fig:2}
\end{center}
\end{figure}

\begin{figure}
\begin{center}
\includegraphics[clip,width=0.85\textwidth]{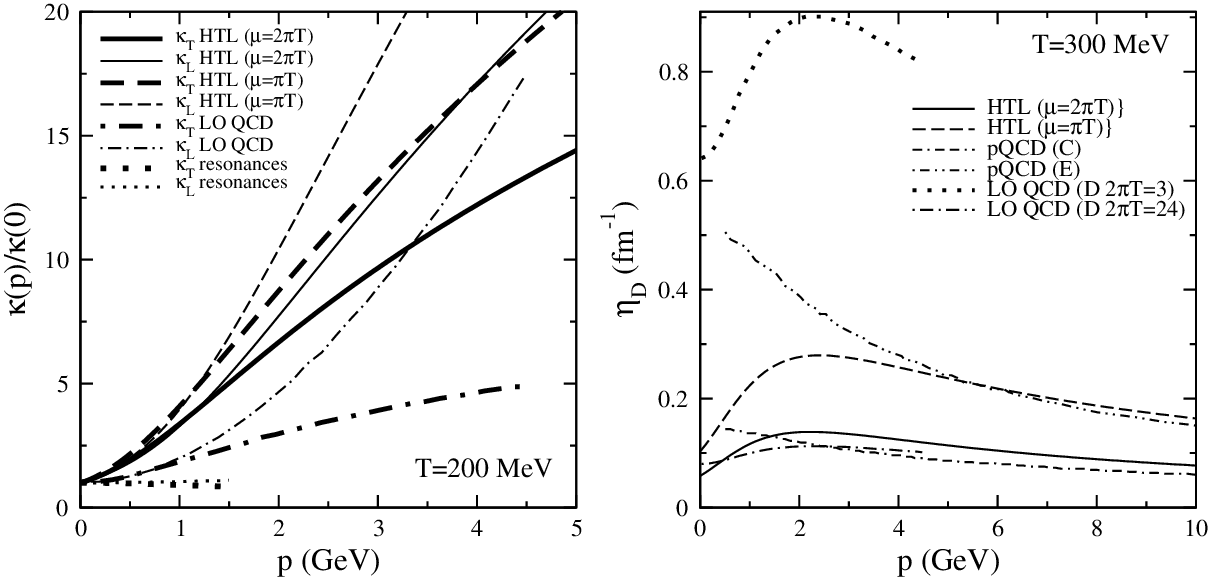}
\caption{Left panel: $\kappa_{T,L}(p)/\kappa_{T,L}(0)$ as a function of
  momentum for charm quarks ($m_c=1.5$~GeV) at $T=200$~MeV in our model (HTL),
  in the the QCD kinetic calculation of Ref.~\protect\cite{tea} (LO~QCD) and in
  the resonance model of Ref.~\cite{rapp1}.
  Right panel: the drag coefficient $\eta_D(p)$ as a function of momentum for 
  charm quarks ($m_c=1.5$~GeV) at $T=300$~MeV in our model (HTL), in the
  calculations of Refs.~\cite{Gos1,Gos3} (pQCD) and \cite{tea} (LO~QCD).}
\label{newfigs}
\end{center}
\end{figure}

In Fig.~\ref{fig:1} we give, for a charm quark and for different temperatures,
the coefficients $\kappa_T(p)$ and $\kappa_L(p)$ which are derived from our HTL
calculation. We also display, for comparison, the corresponding AdS/CFT
result. As expected, the latter model provides stronger coefficients and also a 
much steeper momentum dependence of $\kappa_L(p)$. Notice that, for large enough temperatures, $\gamma_c<1$, so that the
condition $\gamma<\gamma_c$ cannot be satisfied. 
In Fig.~\ref{fig:2} the same quantities for the case of a bottom quark are shown.

In Fig.~\ref{newfigs} we compare our results to a few of the calculations for
the transport coefficients available in the literature.
In the left panel we display the ratios $\kappa_{T,L}(p)/\kappa_{T,L}(0)$ at
$T=200$~MeV in our HTL model together with the QCD kinetic calculation of
Ref.~\cite{tea} (LO~QCD) and the resonance model of Ref.~\cite{rapp1}.
The momentum dependence of the longitudinal coefficient has on the whole a
similar trend in the HTL and LO~QCD models, whereas in the latter the
transverse coefficient displays a less pronounced momentum dependence. On the
other hand, the resonance model provides coefficients that are essentially flat
in the available range of momenta.

In the right panel of Fig.~\ref{newfigs} we compare the drag coefficient
$\eta_D$ as obtained in the HTL model at $T=300$~MeV to the LO~QCD model of
Ref.~\cite{tea}. In that model the value of $\eta_D(0)$ is parameterized
through the diffusion constant in space $D$, which is treated as a free
parameter, and in the figure we display two choices for $D$ in the range
considered in Ref.~\cite{tea}. As one can see, the momentum dependence is
similar to the one of the HTL model, the absolute normalization covering a
wider range of values.

In the same panel we also report the results obtained in Refs.~\cite{Gos1,Gos3}, 
where an emended perturbative QCD (pQCD ) calculation is employed by
introducing a smaller infrared regulator and a suitably chosen running coupling
constant. In the figure the labels (C) and (E) refer to two different choices 
of parameters (see Table~I of Ref.~\cite{Gos1}).
The coefficient $\eta_D$ is smoothly decreasing with $p$ in this model, whereas
in the HTL model (and also in the LO~QCD one) it displays a maximum.
At momenta $p\gtrsim3$~GeV the magnitude of $\eta_D$ is similar in the HTL and
pQCD calculations, the latter, on the other hand, being definetily larger at
small momenta. Also the resonance model of Ref.~\cite{rapp1} displays a smoothly
decreasing behavior with momentum.

\begin{figure*}
\begin{center}
\includegraphics[clip,width=0.85\textwidth]{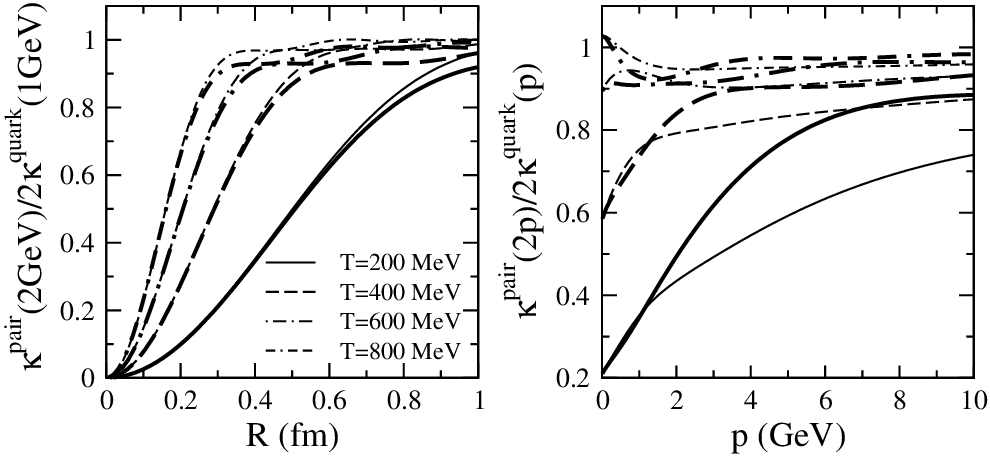}
\caption{The charmonium transport coefficients, normalized to the result for two
  independent $c$-quarks. Left panel: as a function of the $Q\overline{Q}$
  separation, for a pair propagating with momentum
  $p_{c\bar{c}}=2p_c=2$ GeV. Right panel: as a function of the quark 
  momentum, for a pair of size $R=0.4$ fm. Thick lines refer to $\kappa_T$,
  thin lines to $\kappa_L$. Various temperatures are considered.}  
\label{fig:3}
\end{center}
\end{figure*}

\begin{figure*}
\begin{center}
\includegraphics[clip,width=0.85\textwidth]{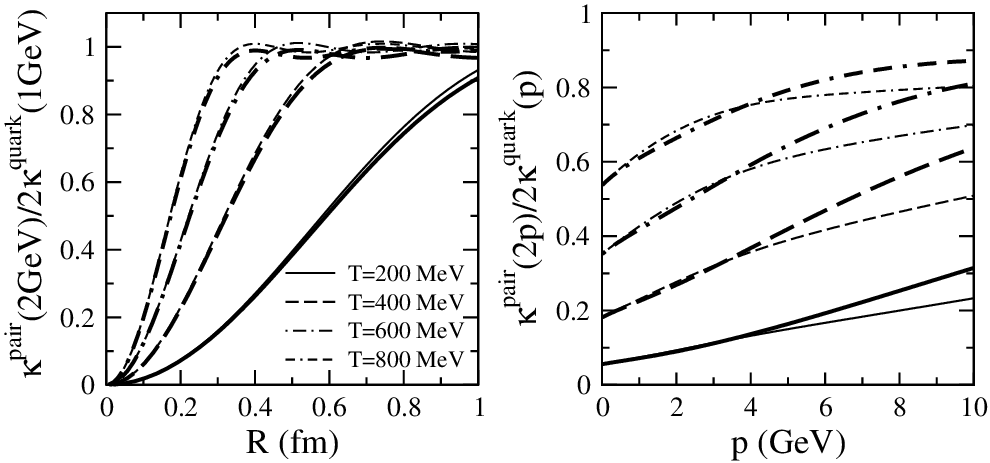}
\caption{The bottomonium transport coefficients, normalized to the result for
  two independent $b$-quarks. Left panel: as a function of the
  $Q\overline{Q}$ separation, for a pair propagating with momentum
  $p_{b\bar{b}}=2p_b=2$ GeV. Right panel: as a function of the quark
  momentum, for a pair of size $R=0.2$ fm. Thick lines refer to $\kappa_T$,
  thin lines to $\kappa_L$. Various temperatures are considered.}  
\label{fig:4}
\end{center}
\end{figure*}

We then consider, in Figs.~\ref{fig:3} and \ref{fig:4}, the case of a
$Q\overline{Q}$ pair crossing the QGP. The quarkonium is modeled as a pair of heavy quarks, which propagates in the medium with a
fixed separation and with a momentum given by the sum of the two individual
quarks. Any binding effect is neglected, hence one has simply 
\begin{eqnarray}
  P_\Phi &=& P_Q+P_{\overline{Q}}=(E_p^Q,\bm{p})+(E_p^{\overline{Q}},\bm{p})
  =(2E_p^Q,2\bm{p})
    \nonumber \\
  &\equiv&(E_{2p}^{\Phi},2\bm{p}),
\end{eqnarray}
with $M_\Phi\equiv2M_c^\text{eff}\approx 2\cdot 1.5$ GeV in the case of
$J/\psi$ and $M_\Phi\equiv2M_b^\text{eff}\approx 2\cdot 4.7$ GeV in the
case of a $\Upsilon$ meson.

In spite of its simplicity such a model is able to display how --- due to
the destructive interference effect encoded in the
$[1-\cos(\bm{q}\cdot\bm{r})]$ factor of Eq.~(\ref{eq:rate_dip1}) ---
quarkonium states of small size propagating in the QGP would suffer much less
rescatterings with respect to two uncorrelated heavy quarks, as it appears from
the left panels of Figs.~\ref{fig:3} and \ref{fig:4}. In particular, since the
relevant quantity is the ratio between the average separation $R$ of the heavy
quarks and the Debye radius $R_D\equiv m_D^{-1}$, for a given dipole size the
suppression is more dramatic at small temperatures, where the Debye radius is
larger, implying a smaller value of $R/R_D$.

In the right panels we still consider the quarkonium transport coefficients,
but as a function of the momentum $p$ of the single ``constituent'' heavy
quark. We take $R=0.4$ fm in Fig.~\ref{fig:3} and $R=0.2$ fm in
Fig.~\ref{fig:4}, corresponding to the typical mean square radius of the
charmonium and bottomonium ground states, respectively. 

We now make use of the above findings for $\kappa_{T/L}(p)$ to follow the
relaxation to thermal equilibrium of a large sample of charm or bottom quarks. 
We model such a process with the following Langevin equation:
\beq\label{eq:lange_r_d0}
\frac{dp^i}{dt}=-\eta_D(p)p^i+\xi^i(t).
\eeq
As in the non-relativistic case introduced in Sec.~\ref{sec:langevin}, the
resulting stochastic evolution of the heavy-quark momenta is
determined by the noise correlation function 
\beq
\label{eq:noise0}
  \langle\xi^i(t)\xi^j(t')\rangle = \delta(t-t')  
  \left[\kappa_L(p)\hat{p}^i\hat{p}^j+
  \kappa_T(p)(\delta^{ij}-\hat{p}^i\hat{p}^j)\right],
\eeq
which is expressed in terms of the transverse and longitudinal transport
coefficients previously evaluated. 
In Eq.~(\ref{eq:lange_r_d0}) the drag coefficients $\eta_D(p)$ is fixed in
order to ensure the approach to equilibrium. An extended discussion of the
relativistic Langevin equation, of its discretization and of the algorithm
employed in its numerical implementation is reported in
Appendix~\ref{sec:app_b}.
We mention here that other choices can be found in the literature. Some authors, for instance, performed a first-principle calculation for $\eta_D$~\cite{hira}, fixing the remaining coefficients by requiring the proper equilibrium limit. Our choice is motivated both by the quest for consistency (since the same microscopic calculation provides both $\kappa_T$ and $\kappa_L$) and by the importance of distinguishing between the transverse and longitudinal momentum broadening of a fast particle (at variance with what done in~\cite{hira}). Finally, we observe that, in any case, in dealing with a stochastic differential equation, the -- momentum-dependent -- friction coefficient receives a correction depending on the adopted discretization scheme, so that it appears more natural to fix the latter ``by hand''. We point out, however, that the HTL calculation, at least at LLA, provides consistent results for $\eta_D$ and $\kappa_L$.

\begin{figure*}
\begin{center}
\includegraphics[clip,width=\textwidth]{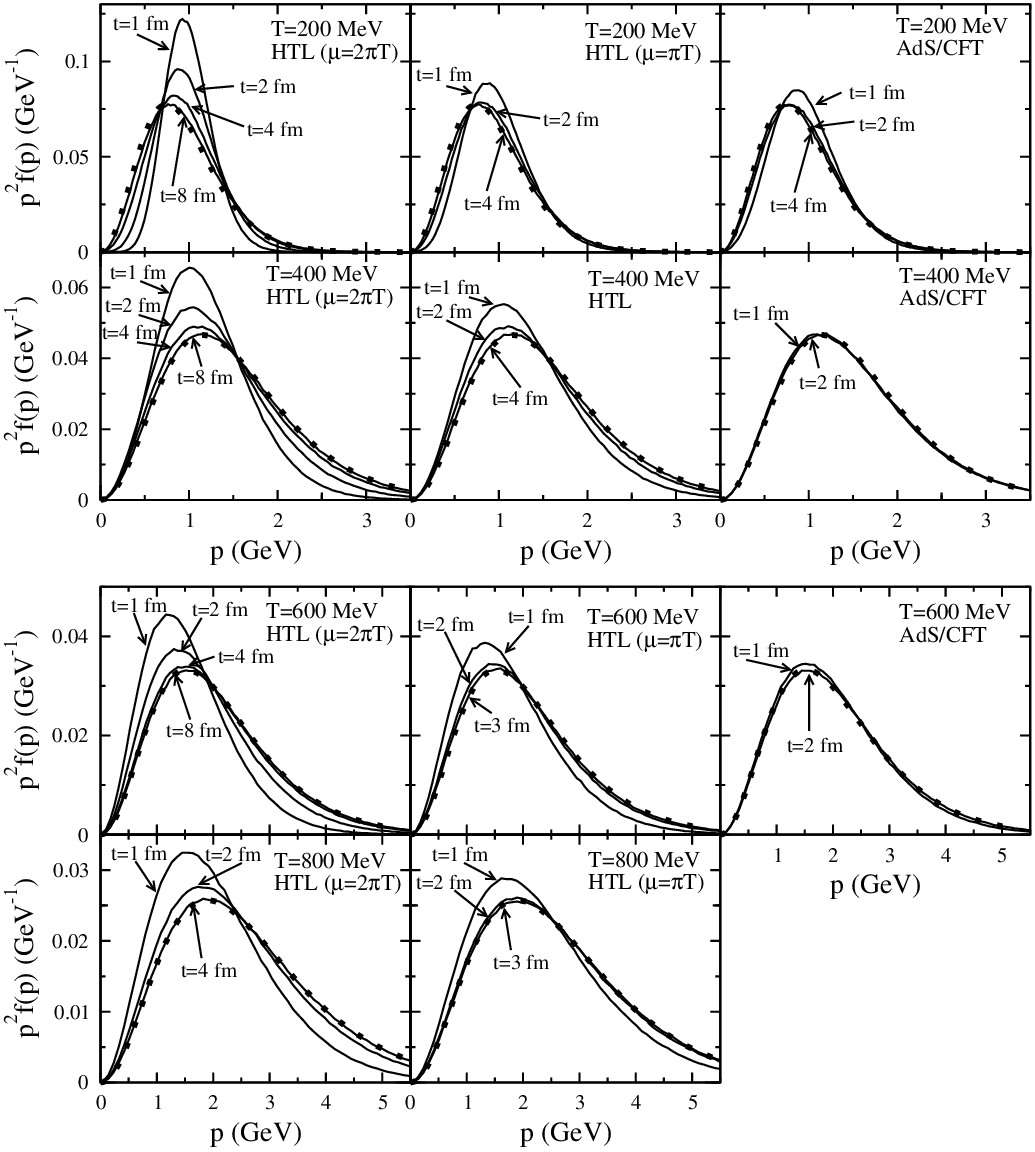}
\caption{The momentum distribution at different times $t$ of charm quarks with 
  $M=1.2$ GeV in a QGP for a range of temperatures. The left and central panels 
  display the HTL results ($\mu=2\pi T$ and $\mu=\pi T$, respectively),
  while the right panels refer to the AdS/CFT calculation. For the initial
  momentum we take $p_0=1$ GeV. Square-dots represent the Maxwell-J\"uttner
  equilibrium distributions.}  
\label{fig:5}
\end{center}
\end{figure*}

\begin{figure*}
\begin{center}
\includegraphics[clip,width=\textwidth]{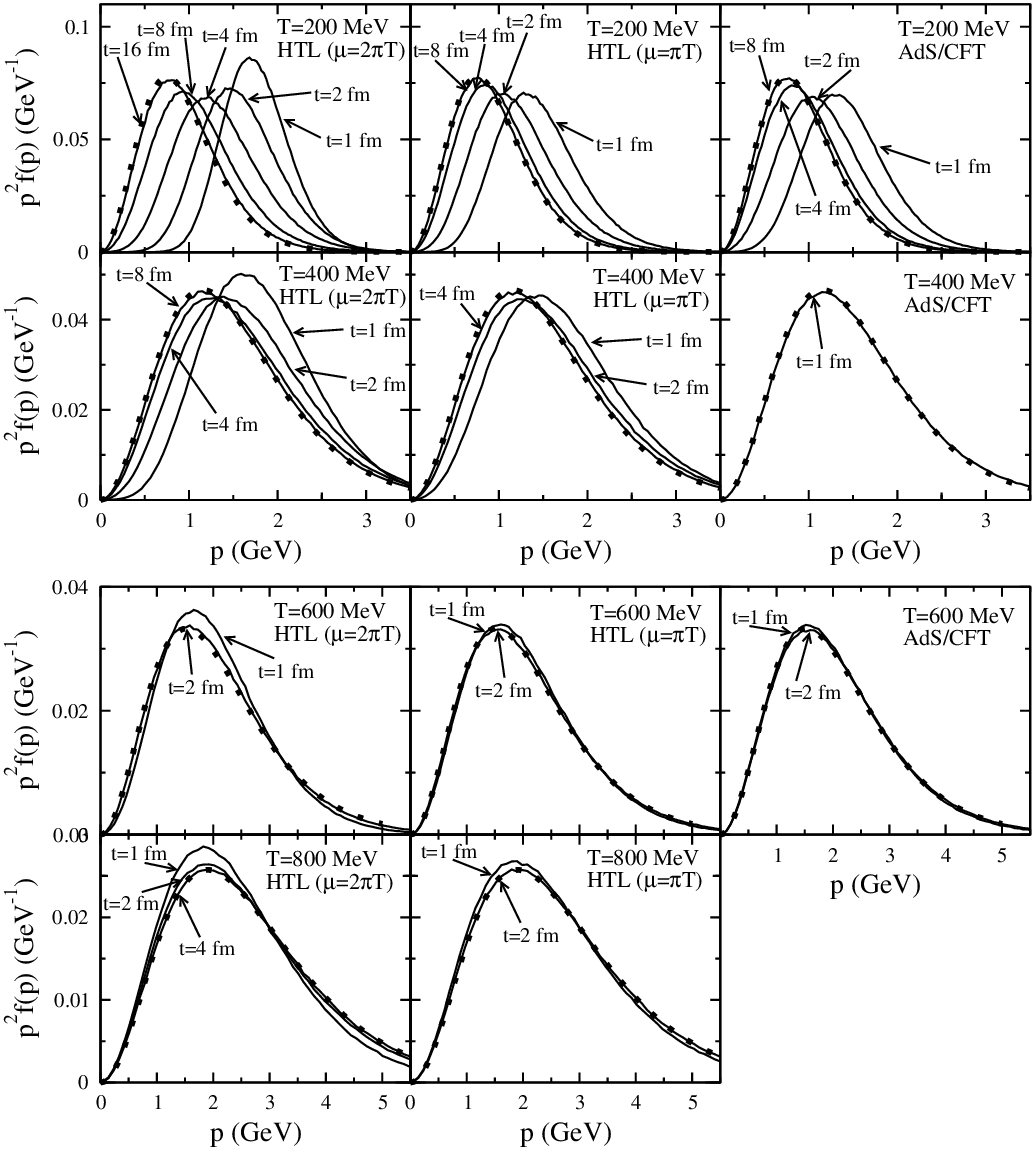}
\caption{As in Fig.~\ref{fig:5}, but for an initial momentum $p_0=2$ GeV.} 
\label{fig:6}
\end{center}
\end{figure*}

We start with the case of a sample of $2\cdot10^6$ charm quarks, with
an initial momentum $p_0$ representative of their typical initial $p_T$ in
nucleus-nucleus collisions. 
The parametrization of their spectrum used in Ref.~\cite{tea}, referring to RHIC
conditions, gives, for instance, $\overline{p}_T=1.23$ GeV. 
In the case of Pb-Pb collisions at LHC (at $\sqrt{s_{NN}}=5.5$ TeV), one
gets, instead, $\overline{p}_T=2.18$ GeV \cite{ppr} for charm
quarks generated by PYTHIA with parameters tuned to reproduce the results by
Mangano {et al.} \cite{MNR}. 
The momentum distributions resulting from the Langevin evolution, for different
values of $T$ and $p_0$, are presented in Figs.~\ref{fig:5} and \ref{fig:6},
for $c$-quarks at increasing values of time.  
For any value of $T$, for large enough times, the charm momenta turn out to be
described by a relativistic Maxwell-J\"uttner distribution 
\beq\label{eq:MJ}
f_\text{MJ}(p)\equiv\frac{e^{-E_p/T}}{4\pi M^2T K_2(M/T)},\;
\text{with}\; \int d^3p f_\text{MJ}(p)=1,
\eeq
where $K_2$ is a modified Bessel function. We notice, however, that the approach
to equilibrium is much faster for larger values of the temperature, due to the
huge increase of the coefficients $\kappa_{T/L}(p)$ with $T$ displayed in
Fig.~\ref{fig:0}. 

\begin{figure*}
\begin{center}
\includegraphics[clip,width=\textwidth]{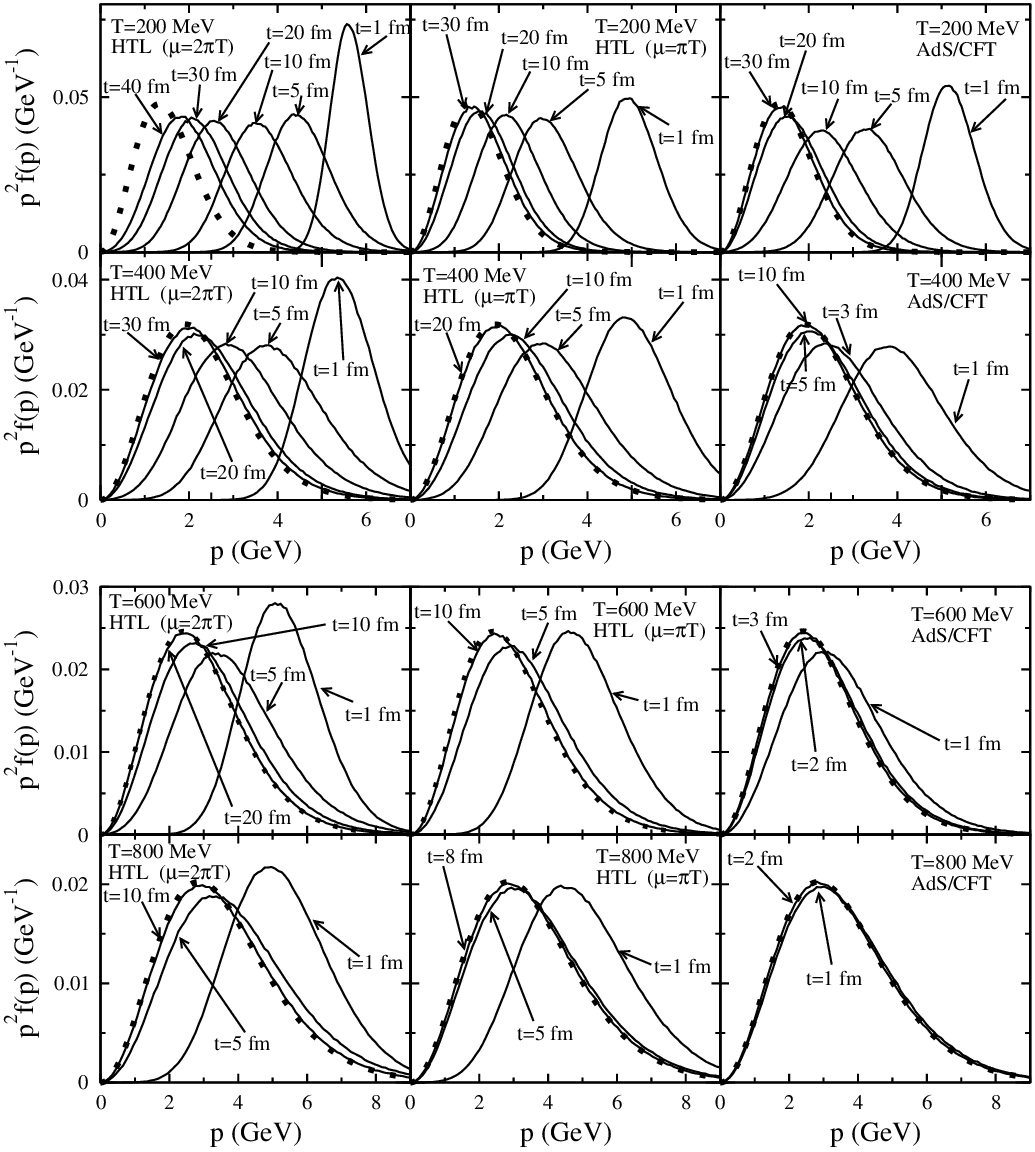}
\caption{As in Fig.~\ref{fig:5}, but for a sample of bottom quarks of
  mass $M=4.2$ GeV and initial momentum $p_0=6$ GeV.} 
\label{fig:7}
\end{center}
\end{figure*}

\begin{figure}
\begin{center}
\includegraphics[clip,width=0.6\textwidth]{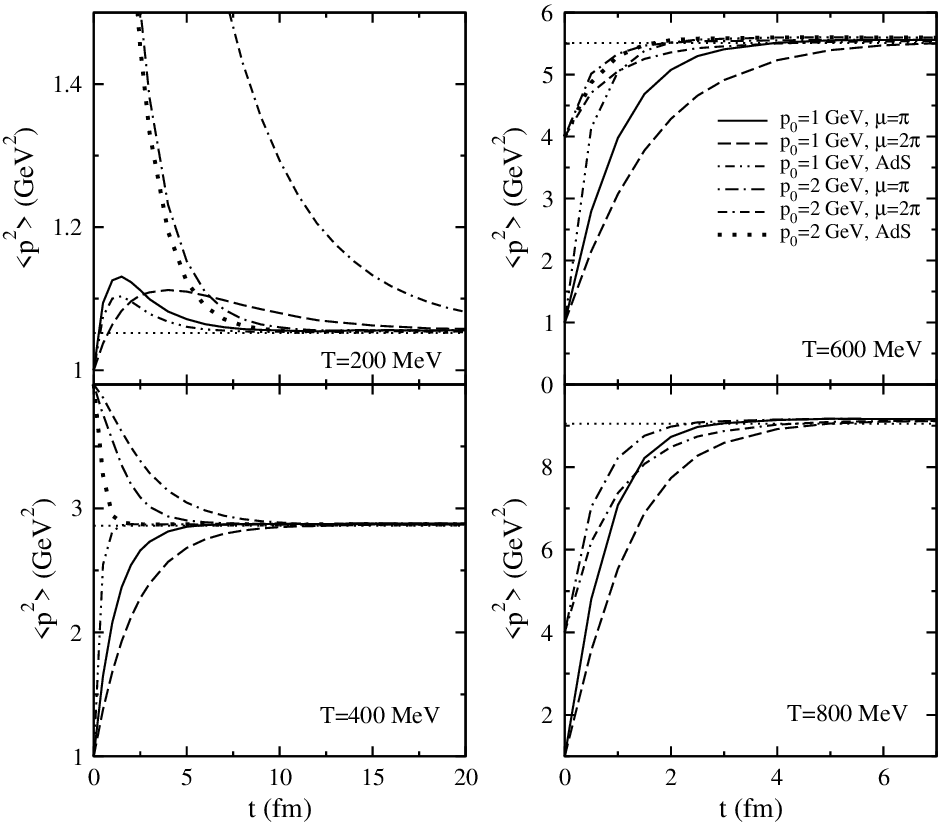}
\caption{The approach of $\langle p^2\rangle$ to the equilibrium value
  predicted by the Maxwell-J\"uttner distribution in the case of charm quarks
  at various temperatures.} 
\label{fig:8}
\end{center}
\end{figure}

\begin{figure}
\begin{center}
\includegraphics[clip,width=0.6\textwidth]{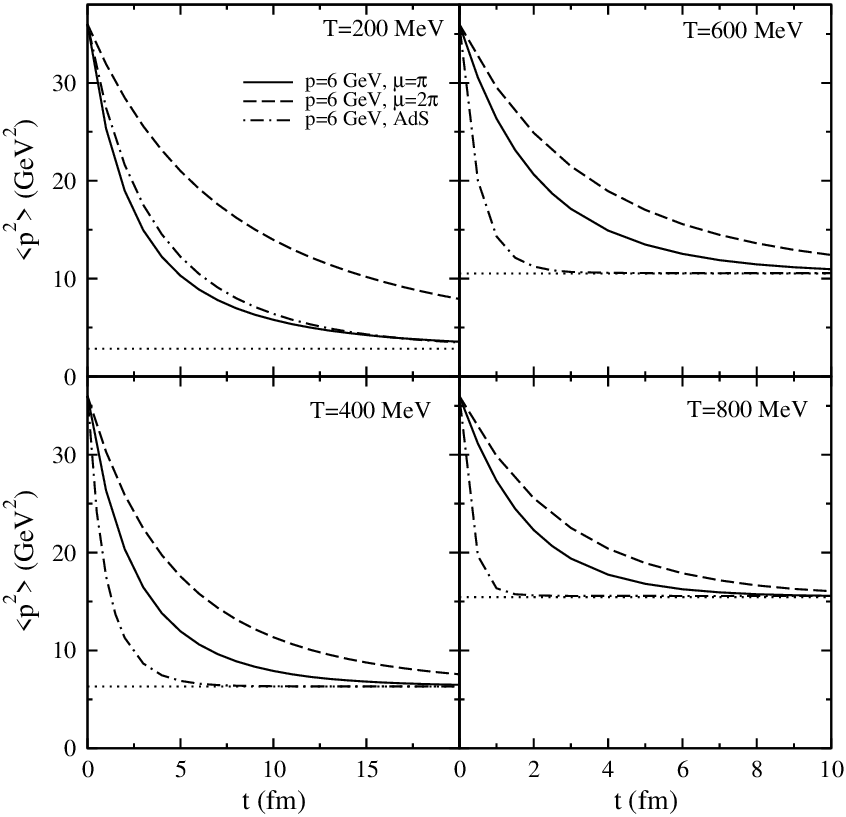}
\caption{As in Fig.~\ref{fig:8}, but in the case of bottom quarks.}
\label{fig:9}
\end{center}
\end{figure}

For comparison we also give, in the right panels, the corresponding results
arising from the AdS/CFT estimate of the transport coefficients entering into
the Langevin equation. 
Such a strongly-coupled scenario gives rise to an extremely fast relaxation
toward equilibrium. 
Notice that the case $T=800$ MeV is out of the range of validity of
the present available methods to compute $\kappa_{T/L}(p)$ within the SYM
framework.

We then consider, for the same range of temperatures, the Langevin dynamics of
an ensemble of $b$-quarks with mass $M=4.2$ GeV, initialized with a
momentum $p_0=6$ GeV, chosen of the order of the average $p_T$ of the
spectrum generated by PYTHIA, with the parameters given in
Ref.~\cite{ppr}. Notably this corresponds to a much harder momentum than
the equilibrium value. This fact, together with the larger mass of the bottom
quark, makes the approach to equilibrium much slower with respect to the case
of charm. 
Our results for the momentum distribution function are displayed in
Fig.~\ref{fig:7}. As it can be seen the HTL calculation provides relaxation
times too large to be of interest for the realistic case of an expanding (and
cooling) fireball. 

On the contrary the AdS/CFT scenario supports a quite rapid thermalization
also for $b$-quarks, the latter being almost immediate at the largest
temperatures examined.

Finally, in order to get a feeling of the rapidity of the thermalization
process in the different cases considered here (which can be not so easy to
grasp just by looking at the distributions), we give in Figs.~\ref{fig:8} and
\ref{fig:9} the evolution of the mean squared momentum $\langle p^2\rangle$ to
its equilibrium value predicted by the Maxwell-J\"uttner distribution of
Eq.~(\ref{eq:MJ}). The difference between the two models (HTL and SYM) appears
particularly evident in the case of $b$-quarks, which could be an interesting
probe to study at LHC in order to discriminate between different scenarios. 

\section{Discussion and conclusions}\label{sec:concl}

The two main issues addressed in this paper are the evaluation of transport
coefficients for heavy quarks in the QGP and the study of the resulting
Langevin evolution of their momenta. 
We showed that important qualitative and quantitative (though affected by
some systematic uncertainty) information on the energy-loss and
momentum-broadening of heavy quarks can be extracted from an approach (based on
the HTL approximation) developed for a different purpose, namely for the
definition of an effective in-medium potential for a pair of static quarks
\cite{hq0,hq1}. 

Concerning the transport coefficients, both $\kappa_T$ and $\kappa_L$ were
found to display a sizable growth with the momentum, $\kappa_L$ increasing
faster than $\kappa_T$.

The approach developed in this work has been also useful for the study of
transport properties of quarkonia.
$Q\overline{Q}$ states of small size, behaving as an almost neutral object, turned out to suffer less rescatterings. This could be of
phenomenological interest in order to discriminate between different production/suppression scenarios. Analysis of momentum spectra would, in fact, help to understand
whether the observed quarkonia in the final state arise from the recombination
of uncorrelated heavy quarks or from their primordial production, after
crossing unaffected the deconfined environment. 

We then addressed the Langevin evolution of a large sample of heavy quarks
(charm and bottom) --- starting from an initial momentum chosen to be 
representative of the typical ones in nucleus-nucleus collisions --- and 
following their approach to thermalization. We also tried to give some basic
discussion of the  essential aspects of the relativistic Langevin equation as
an effective model to describe the relaxation of an ensemble of heavy particles
toward equilibrium.

Our numerical findings allow to draw some general considerations.

We first notice that at the largest temperatures explored ($T=600$ and
$T=800$ MeV) the thermalization of $c$-quarks predicted by our
(weak-coupling) HTL calculation is quite fast. Such an occurrence could be of
phenomenological interest for LHC, where one expects quite large temperatures
to be achieved in the first instants after the collision. 
On the contrary the relaxation of $b$-quark spectra toward equilibrium arising
from the present HTL calculation is always too slow to allow
them to follow the flow of the medium in the realistic case of an expanding
fireball.

For what concerns the strongly-coupled AdS/CFT scenario, $c$-quarks are found
to thermalize almost immediately, in spite of our conservative choice of
freezing the values of $\kappa_{T/L}$ for $\gamma<\gamma_c$. Interestingly,
also $b$-quarks approach equilibrium quite fast (at the largest temperatures),
making possible for them to inherit at least part of the flow of the medium
produced in heavy-ion collisions. Elliptic-flow is in fact expected to develop
in the very initial stage of the fireball evolution, when the medium is
extremely hot, entailing large values of $\kappa_{T/L}$ (which, we remind, grow
as $T^3$). 

At the large temperatures realized at LHC, $b$-quarks appear then as a promising
probe to discriminate between the weakly and strongly-coupled scenarios. A
sizable flow of $b$-quarks could not be compatible with any weak-coupling
perturbative calculation. Measurement of the elliptic-flow and of the quenching
of $p_T$ spectra of single-electrons (or positrons) arising from the
semi-leptonic decays of $B$-mesons (which will be possible at LHC) could then
shed light on the transport properties of the QGP at the future accessible
regime of temperatures.

In future work we plan to employ the approach developed here, whose numerical
implementation turns out to be quite inexpensive, in order to address the 
realistic case of a fireball displaying longitudinal and transverse
expansion.
Supplementing our study with a more microscopic kinetic calculation of the hard-scattering contribution to the transport coefficients is also object of our current investigation.

\section*{Acknowledgments}
We are grateful to F. Prino for fruitful discussions and for providing us with
the heavy-quark production spectra expected at LHC.

\appendix

\section{The HTL propagators and spectral functions}
\label{app:a}

We give here the explicit expressions for the HTL gluon (photon in QED) 
propagators in the Coulomb gauge, together with their spectral functions, which
are employed throughout the text. Further details can be found, e.g., in
Ref.~\cite{pr}.

The longitudinal and transverse \emph{analytical} (off the real-energy axis)
propagators read, respectively,
\beqa
\Delta_L(q^0,q)&=&\frac{-1}{q^2+\Pi_L(x)},\nonumber\\
\Delta_T(q^0,q)&=&\frac{-1}{(q^0)^2-q^2-\Pi_T(x)},
\eeqa
where
\beqa
\Pi_L(x)&=&m_D^2\left[1-Q(x)\right],\nonumber\\
\Pi_T(x)&=&\frac{m_D^2}{2}\left[x^2+(1-x^2)Q(x)\right]\quad \text{and}
\nonumber\\
Q(x)&\equiv&\frac{x}{2}\ln\frac{x+1}{x-1},
\eeqa
being $x\equiv q^0/q$.\\
The corresponding \emph{retarded} propagators are obtained by setting
$q^0=\omega+i\eta$ in the above expressions, namely:
\beq
D_{L/T}^R(\omega,q)\equiv\Delta_{L/T}(\omega+i\eta,q).
\eeq
The HTL gluon (photon) spectral function follows immediately from
the definition:
\beq
\rho_{L/T}(\omega,q)\equiv 2 \text{Im}D_{L/T}^R(\omega,q).
\eeq
One gets (for space-like momenta):
\beqa
&&\rho_L(\omega,q)=\frac{\displaystyle{2\pi m_D^2\frac{\omega}{2q}}}
{\displaystyle{\left[q^2+m_D^2\left(1-\frac{\omega}{2q}
\ln\left|\frac{\omega+q}{\omega-q}\right|\right)\right]^2
+\left(\pi m_D^2\frac{\omega}{2q}\right)^2}}, \\
&&\rho_T(\omega,q)= 
\frac{\displaystyle{\pi m_D^2\frac{\omega(q^2-\omega^2)}
{2q^3}}}
{\displaystyle{\left[\omega^2-q^2-\frac{m_D^2}{2}\frac{\omega^2}{q^2}
\left(1-\frac{\omega^2-q^2}{2\omega q}
\ln\left|\frac{\omega+q}{\omega-q}\right|\right)\right]^2
+\left[\frac{\pi}{2}m_D^2\frac{\omega(\omega^2-q^2)}{2q^3}\right]^2}}.\nonumber
\eeqa
Their expressions at LLA are obtained by neglecting the self-energy corrections
in the denominators and are given by:
\beqa
\rho_L(\omega,q)&\underset{\text{LLA}}{\sim}&
\frac{\displaystyle{2\pi m_D^2\frac{\omega}{2q}}}
{q^4},\nonumber\\
\rho_T(\omega,q)&\underset{\text{LLA}}{\sim}&
\frac{\displaystyle{\pi m_D^2\frac{\omega}
{2q}}}
{q^2(q^2-\omega^2)}.\label{eq:LLAsp}
\eeqa
This implies that the infrared divergences are no longer screened by
the Debye mass, and must be regulated by hand by setting a lower cutoff of
order $m_D$ in the integration over the three-momentum of the exchanged gluons
(photons).

When dealing with the in-medium heavy quark propagation, within the eikonal
approximation, one needs to consider the exponentiation of the 
\emph{real-time} gauge-field propagator; the latter is given by
\beqa
  D_L(\omega,q)&=&-\frac{1}{q^2}
    +\int_{-\infty}^{+\infty}\frac{dq^0}{2\pi}
    \frac{\rho_L(q^0,q)}{q^0-(\omega+i\eta)}+i\rho_L(\omega,q)N(\omega)
    \nonumber\\
  D_T(\omega,q)&=&\int_{-\infty}^{+\infty}\frac{dq^0}{2\pi}
    \frac{\rho_T(q^0,q)}{q^0-(\omega+i\eta)}+i\rho_T(\omega,q)N(\omega),
\eeqa
where $N(\omega)$ is the customary Bose distribution. The above expressions are
related to the respective retarded propagators as follows: 
{\setlength\arraycolsep{1pt}
\beqa
D_{L/T}(\omega,q)&=&D_{L/T}^R(\omega,q)+i\rho_{L/T}(\omega,q)N(\omega)
\\
{}&=& \text{Re}\,D_{L/T}^R(\omega,q)+i\left(N(\omega)+\frac{1}{2}\right)
\rho_{L/T}(\omega,q).\nonumber
\eeqa}

\section{The relativistic Langevin equation}
\label{sec:app_b}

In Sec.~\ref{sec:langevin} we discussed the non-relativistic Langevin dynamics
of a heavy-quark in a hot plasma. This allowed us to 
take advantage of well known results like the Einstein relation given in 
Eq.~(\ref{eq:einstein_nr}), which provides a link between the friction and the
momentum-diffusion coefficients and can be seen as an example of the
fluctuation-dissipation theorem. 

The generalization of the Langevin equation to the relativistic case is far 
from trivial. In particular one can no longer ignore the momentum dependence of
the transport coefficients, as it is usually done in non-relativistic studies;
in turn, this introduces in the definition of the 
friction term a dependence on the adopted discretization scheme.

From the mathematical point of view the relativistic Langevin equation belongs
to the class of \emph{stochastic differential equations} with multiplicative
noise, which represents by itself an active field of investigation
\cite{stocha}.  
Although the traditional Langevin treatment of non-equilibrium dynamics is
textbook material, the case of state-dependent (momentum dependent in the
present case) transport coefficients has been only rarely considered in the
literature \cite{tea,rapp2,rapp3,hira,arn1,arn2,lau} and from quite different
perspectives. 
Hence we provide in this Appendix a self-contained discussion of the essential
aspects of the relativistic Langevin equation. We mainly follow the 
approach presented in Ref.~\cite{lau}, which we found particularly clear.

For simplicity, let us consider the one-dimensional case, which is sufficient to
illustrate the relevant conceptual points. 
The generalization to the realistic $d$-dimensional case will follow
immediately. 
Also in the relativistic case one postulates that the momentum of
an external particle placed in a hot medium evolves according to the equation 
\beq\label{eq:lange_r}
\frac{dp}{dt}=-\eta_D(p)p+\xi(t),
\eeq
where the right hand side is given again by the sum of a friction force and a
term related to the random momentum kicks received from the medium particles 
\beq\label{eq:noise_r}
\langle\xi(t)\xi(t')\rangle=\kappa(p)\delta(t-t'),
\eeq
still assumed to be uncorrelated. This is expressed by the delta
function, leading to a flat power spectrum referred to as \emph{white noise};
however the strength $\kappa$ of the noise in Eq.~(\ref{eq:noise_r}) depends,
in the general case, on the momentum of the brownian particle
(\emph{multiplicative noise}). 
One can factor out the momentum dependence by defining
$g(p)\equiv\sqrt{\kappa(p)}$, so that Eq.~(\ref{eq:lange_r}) becomes
\beq\label{eq:lange_r2}
\frac{dp}{dt}=-\eta_D(p)p+g(p)\eta(t),
\eeq
with
\beq\label{eq:corr_eta}
\langle\eta(t)\eta(t')\rangle=\delta(t-t').
\eeq
We anticipate that the expression of the friction coefficient $\eta_D(p)$
will depend on the discretization scheme, namely on the value of $p$ at which
in the discretized equation (employed in the numerical simulations)
\beq\label{eq:lange_r_dis}
\Delta p\equiv p(t+\Delta t)- p(t)=-[\eta_D(p)p]\Delta t +g(p)\eta(t)\Delta t
\eeq
the terms in the right hand side must be evaluated. In the following we wish to
clarify this point.

The Langevin equation can be written in the general form
\beq\label{eq:lange_r_gen}
\frac{dp}{dt}=f(p)+g(p)\eta(t),
\eeq
where we can identify
\begin{eqnarray}
  f(p)&\equiv&-\eta_D(p)p \equiv-\eta_D^{(0)}(p)p+f_1(p) \nonumber\\ 
  &=& -\left[\eta_D^{(0)}(p) - \frac{f_1(p)}{p}\right] p.
\end{eqnarray}
In the above the friction coefficient has been written as the sum of a leading
term, linked to the momentum diffusion coefficient $\kappa(p)$ by a relation
analogous to the one occurring in the non-relativistic case, and a
subleading (in $T/E$) correction which will depend on the discretization
procedure. The ambiguity will be eliminated by requiring that in the continuum
limit all the different discretization schemes yield to the same
Fokker-Planck equation for the momentum distribution, which admits the
equilibrium distribution $\exp{(-E_p/T)}$ as a steady solution.

We start by integrating Eq.~(\ref{eq:lange_r_gen}), which leads to the formal
expression
\beq
p(t+\Delta t)- p(t)=\int_{t}^{t+\Delta t}ds[f(p(s))+g(p(s))\eta(s)].
\eeq
However the random noise term $\eta(s)$ makes the value of the integral
ambiguous and a recipe has to be given to evaluate the right hand side of the 
above equation.
One can overcome this difficulty by considering a whole family of different
discretizations, labeled by a parameter $\alpha\in[0,1]$, such that
\beq\label{eq:discret}
\Delta p=f[p(t)+\alpha\Delta p]\Delta t+g[p(t)+\alpha\Delta p]
\int_{t}^{t+\Delta t}ds
\,\eta(s).
\eeq
Note that the momentum increment $\Delta p$ arises from the sum of a
deterministic friction term of order $\Delta t$ and of a random term of order
$\sqrt{\Delta t}$. After defining $p_0\equiv p(t)$, the first term in the
above, being already of order $\Delta t$, can be evaluated at $p_0$, while for
the second term it is useful to expand
\beq
g[p_0+\alpha\Delta p]=g(p_0)+g'(p_0)\,\alpha\Delta p+\dots,
\eeq
where the last term, multiplying the noise-integral, provides also a
contribution of order $\Delta t$. Taking advantage of the fact that
$\langle\eta(s)\rangle=0$ and of Eq.~(\ref{eq:corr_eta}) one immediately
gets:
\beq\label{eq:1mom}
\langle\Delta p\rangle=f(p_0)\Delta t+\alpha g(p_0)g'(p_0)\Delta t
\eeq
and
\beq\label{eq:2mom}
\langle(\Delta p)^2\rangle=g^2(p_0)\Delta t.
\eeq
We now consider the link between the Langevin equation and the Fokker-Planck
equation for the momentum distribution $P(p,t)$ of the brownian particles.
For the latter the following equation holds:
\beq\label{eq:condprob}
P(p,t+\Delta t)=\int_{-\infty}^{+\infty} dp_0 P(p,t+\Delta t|p_0,t)
P(p_0,t),
\eeq
where $P(p,t+\Delta t|p_0,t)$ represents the \emph{conditional probability}
that a particle with momentum $p_0$ at time $t$ will be found with momentum
$p$ at time $t+\Delta t$.
One can identify such a conditional probability with the following expectation
value over the ensemble of brownian particles: 
\beqa
  P(p,t+\Delta t|p_0,t)&\equiv&
    \langle\delta[p-p(t+\Delta t)]\rangle_{p_0,t} \nonumber\\
  &=&\langle\delta[p-p_0-\Delta p]\rangle_{p_0,t},
\eeqa
with $\Delta p$ taken from the Langevin equation (\ref{eq:discret}). One can
then expand up to second order, obtaining:
\beq
P(p,t+\Delta t|p_0,t)=\delta(p-p_0)-\langle\Delta p\rangle\frac{\partial}
{\partial p}\delta(p-p_0)+\frac{1}{2}\langle(\Delta p)^2
\rangle\frac{\partial^2}{\partial p^2}\delta(p-p_0)+\dots
\eeq
After inserting the above expansion into Eq.~(\ref{eq:condprob}) and exploiting
Eqs.~(\ref{eq:1mom}) and (\ref{eq:2mom}), one arrives to the Fokker-Planck 
equation:
\beqa
\frac{\partial}{\partial t}P(p,t)&=&
\frac{\partial}{\partial p}\left[-f(p)-\alpha
g(p)g'(p)+\frac{1}{2}\frac{\partial}{\partial p}g^2(p)\right]P(p,t)\nonumber\\
{}&=&\frac{\partial}{\partial p}\left[\eta_D^{(0)}(p)p-f_1(p) -\alpha
g(p)g'(p)+\frac{1}{2}\frac{\partial}{\partial p}g^2(p)\right]P(p,t)\nonumber\\
{}&=&\frac{\partial}{\partial p}\left[\eta_D^{(0)}(p)p-f_1(p)+\frac{1}{2}
(1-\alpha)\kappa'(p)+\frac{1}{2}\kappa(p)\frac{\partial}{\partial p}\right]
P(p,t). 
\eeqa
Then, the requirement that the above equation is independent of the
discretization scheme (i.e. of $\alpha$) and admits the steady solution
$\exp{(-E_p/T)}$, with $E_p\equiv\sqrt{p^2+M^2}$, allows to fix in an
unambiguous way $\eta_D^{(0)}(p)$ and $f_1(p)$. One gets:
\beqa
\eta_D^{(0)}(p)&=&\frac{\kappa(p)}{2TE_p}\\
f_1(p)&=&\frac{1}{2}(1-\alpha)\partial_p\kappa(p).
\eeqa
Hence the friction coefficient to be employed in the numerical Langevin
simulations depends on the discretization scheme through the parameter $\alpha$ 
and reads:
\beq
\eta_D(p)=\frac{\kappa(p)}{2TE_p}-
\frac{1}{2}(1-\alpha)\frac{\partial_p\kappa(p)}{p}. 
\eeq
Two very popular choices in the literature are $\alpha=0$
(Ito discretization) and $\alpha=1/2$ (Stratonovich discretization).
For our purposes it will result more convenient to keep track of the
dependence on the velocity of the brownian particle $v=p/E_p$ rather then
on its momentum, so that:
\beq
\eta_D(v)=\frac{\kappa(v)}{2TE_p}-\frac{1}{2}(1-\alpha)\frac{1-v^2}{pE_p}
\partial_v\kappa(v).
\eeq
We notice that the Fokker-Planck equation turns out to be completely determined
by the momentum-diffusion coefficient: 
\beq
\frac{\partial}{\partial t}P(p,t)=\frac{\partial}{\partial p}
\left\{ \frac{1}{2}\kappa(p)\left[\frac{p}{T E_p}+\frac{\partial}{\partial p}
\right]P(p,t) \right\}.
\eeq

We now consider the relativistic brownian motion in $d$ space dimensions.
The Langevin equation in this case reads:
\beq\label{eq:lange_r_d}
\frac{dp^i}{dt}=-\eta_D(p)p^i+\xi^i(t),
\eeq
with 
\beq\label{eq:noise1}
\langle\xi^i(t)\xi^j(t')\rangle=b^{ij}(\bm{p})\delta(t-t'),
\eeq
where
\beq\label{eq:noise2}
b^{ij}(\bm{p})\equiv \kappa_L(p)\hat{p}^i\hat{p}^j+\kappa_T(p)
(\delta^{ij}-\hat{p}^i\hat{p}^j).
\eeq
It is also useful to introduce the tensor
\beqa
g^{ij}(\bm{p})&\equiv&\sqrt{\kappa_L(p)}\hat{p}^i\hat{p}^j+\sqrt{\kappa_T(p)}
(\delta^{ij}-\hat{p}^i\hat{p}^j)\nonumber\\
{}&\equiv&
g_L(p)\hat{p}^i\hat{p}^j+g_T(p)(\delta^{ij}-\hat{p}^i\hat{p}^j).
\eeqa
This allows one to factor out the momentum dependence of the noise term in
Eq.~(\ref{eq:lange_r_d}), which becomes
\beq\label{eq:lange_r_d2}
\frac{dp^i}{dt}=-\eta_D(p)p^i+g^{ij}(\bm{p})\eta^i(t),
\eeq
with
\beq
\langle\eta^i(t)\eta^j(t')\rangle=\delta^{ij}\delta(t-t').
\eeq
Eq.~(\ref{eq:lange_r_d2}) can be viewed as a particular case of the generic
stochastic equation
\beq\label{eq:lange_r_dgen}
\frac{dp^i}{dt}=f^i(\bm{p})+g^{ij}(\bm{p})\eta^i(t).
\eeq
One can then repeat the same steps followed in the one-dimensional case.
From the Langevin equation one gets the expectation values:
{\setlength\arraycolsep{1pt}
\beqa
\langle\Delta p^i\rangle&=&f^i(\bm{p}_0)\Delta t+\alpha
\left(\partial_k g^{ij}(\bm{p}_0)\right)g^{kj}(\bm{p}_0)\Delta t\nonumber\\
\langle\Delta p^i\Delta p^j \rangle&=& g^{ik}(\bm{p}_0)g^{jk}(\bm{p}_0)\Delta t
=b^{ij}(\bm{p}_0)\Delta t,
\eeqa}
which lead to the Fokker-Planck equation:
\beqa
\frac{\partial}{\partial t}P(\bm{p},t)&=&\frac{\partial}{\partial p^i}
\left[-f^i(\bm{p})-\alpha\left(\partial_k g^{ij}(\bm{p})\right)g^{kj}(\bm{p})
+\frac{1}{2}\frac{\partial}{\partial p^j}b^{ij}(\bm{p})\right]
P(\bm{p},t) \\
{}&=&\frac{\partial}{\partial p^i}
\left[-f^i(\bm{p})+\frac{1}{2}\partial_j\left(b^{ij}(\bm{p})\right)- 
\alpha\left(\partial_k g^{ij}(\bm{p})\right)g^{kj}(\bm{p})
+\frac{1}{2} b^{ij}(\bm{p})\frac{\partial}{\partial p^j}\right]P(\bm{p},t).\nonumber
\eeqa
Also in the $d$-dimensional case, requiring the Fokker-Planck equation to be
independent on the discretization scheme and to admit a relativistic Maxwell
distribution as a steady solution allows to fix the friction term entering into
the Langevin equation and to relate it to the momentum-diffusion
coefficients. One gets: 
\beq
f^i(\bm{p})=-\frac{1}{2T}b^{ij}(\bm{p})\frac{\partial E_p}{\partial p^j}
+\frac{1}{2}\frac{\partial  b^{ij}(\bm{p})}{\partial p^j} \\
-\alpha\left(\partial_k g^{ij}(\bm{p})\right)g^{kj}(\bm{p}),
\eeq
which leads to
\beqa
  f^i(\bm{p}) &=& -\frac{\kappa_L(p)}{2TE}p^i + \frac{1}{2}\left[\partial_p\kappa_L(p)+
    \frac{d-1}{p}(\kappa_L(p)-\kappa_T(p))\right]\hat{p}^i \nonumber\\
  &&-\alpha\left[g_L(p)\partial_p g_L(p)+
    \frac{d-1}{p}\,g_T(p)(g_L(p)-g_T(p))\right]\hat{p}^i. 
\eeqa
The friction coefficient to be employed in the Langevin equation can be then
conveniently written (for the sake of simplicity we give it for the
$\alpha=0$ Ito discretization) as:
\beq
  \eta_D^{\text{Ito}}(p)=\frac{\kappa_L(p)}{2TE} 
-\frac{1}{E^2}\left[(1-v^2)\frac{\partial\kappa_L(p)}{\partial v^2}+
\frac{d-1}{2}\,\frac{\kappa_L(p)-\kappa_T(p)}{v^2}\right], 
\eeq
to be compared with the result given in Ref.~\cite{tea}.
The recipe employed to update the heavy-quark momentum is then the following:
\beqa
p_{n+1}^i-p_n^i&=&-\eta_D^{\text{Ito}}(p_n)p_n^i\Delta t+\xi^i(t_n)\Delta
t\\ {}&\equiv&
-\eta_D^{\text{Ito}}(p_n)p_n^i\Delta t+g^{ij}(\bm{p}_n)\zeta^i(t_n)\sqrt{\Delta
  t},\nonumber 
\eeqa
with
\beq\label{eq:update2}
\langle\zeta^i(t_n)\zeta^j(t_m)\rangle=\delta_{m,n}\delta^{i,j}.
\eeq
Hence, at each time-step and for each quark, one has simply to extract $d$
independent random numbers from a gaussian distribution with $\sigma=1$,
as it can be seen from Eq.~(\ref{eq:update2}). 

\section{Estimate of $q_\text{max}$}\label{sec:app_c}

In order to obtain a realistic estimate of $q_\text{max}$ to employ in the
numerical calculations we adopt the following strategy. 
We take, as the upper bound of integration over the exchanged momenta, the
ultraviolet cutoff given in Ref.~\cite{gyu}. The latter, in the limit $E_p\gg
T$ required by our eikonal approach to be reliable, reads: 
\beq\label{eq:match1}
q_\text{max}=\frac{2\langle k\rangle(E+p)}{\sqrt{M^2+2\langle k\rangle(E+p)}}.
\eeq
In the above $\langle k\rangle\sim T$ represents an average momentum of the
thermal particles taking part to the collisions. In Ref.~\cite{gyu} the authors
let it vary from $T$ to $3T$. 
In order to avoid plotting too many curves, our procedure is instead the
following. We consider the non-relativistic limit (i.e. $M\gg T$) and we match
the results for the momentum diffusion coefficient $\kappa$ given by the QCD
kinetic calculation \cite{tea} 
\beqa
3\kappa^{nr}&=&C_F\frac{g^4}{2\pi^3}\int_0^{\infty} k^2 
dk\int_0^{2k}dq\frac{q^3}{(q^2+m_D^2)^2} \\
 && \times\left[
\frac{N_f}{2}\frac{e^{\beta k}}{(e^{\beta k}+1)^2}\left(2-
\frac{q^2}{2k^2}\right) +\frac{N_c}{2}\frac{e^{\beta k}}{(e^{\beta k}-1)^2}
\left(2-\frac{q^2}{k^2}+\frac{q^4}{4k^4}\right)
\right] \nonumber
\eeqa
and by the HTL effective approach, namely
\beq
3\kappa^{nr}=C_F\frac{g^2 m_D^2
  T}{2\pi}\int_0^{q^\text{nr}_\text{max}}\frac{q^3dq}{(q^2+m_D^2)^2}, 
\eeq 
where
\beq
m_D^2=g^2T^2\left(\frac{N_c}{3}+\frac{N_f}{6}\right).
\eeq
We found that, for the case of a coupling running with the temperature, the 
expression 
\beq
q^\text{nr}_\text{max}=3.1Tg^{1/3}(T)
\eeq
provides a good matching, with an agreement at the $1\%$ level over the whole
range of temperatures. Inserting the above expression into the zero momentum
limit of Eq.~(\ref{eq:match1}) allows then to fix $\langle k \rangle$, and
hence to determine the corresponding bound for general values of $p$.

\end{document}